%% file: main.tex
\documentclass[compsoc,conference,a4paper,10pt,times]{IEEEtran}
\IEEEoverridecommandlockouts
\usepackage{cite}
\usepackage{amsmath,amssymb,amsfonts}
\usepackage{algorithmic}
\usepackage{graphicx}

\usepackage[caption=false]{subfig}
\renewcommand{\figurename}{Fig.}
\usepackage{textcomp}
\usepackage{bmpsize}
\usepackage[dvipsnames]{xcolor}
\usepackage{lipsum}
\usepackage{booktabs}
\usepackage[colorlinks=true,linkcolor=blue,breaklinks=True,citecolor=brown,urlcolor=blue]{hyperref}
\def\BibTeX{{\rm B\kern-.05em{\sc i\kern-.025em b}\kern-.08em
    T\kern-.1667em\lower.7ex\hbox{E}\kern-.125emX}}
\usepackage{url} 
\usepackage{array} 
\usepackage{xurl}
\usepackage[most]{tcolorbox}
\input{commands}
\usepackage{enumitem}
\setlist[itemize]{noitemsep, topsep=0pt}
\usepackage{pifont}
\usepackage[sort&compress, numbers]{natbib}
\usepackage{tikz}
\usepackage{xspace} 
\usepackage{multirow}
\usepackage{graphicx}
\usepackage{subfig}
\usepackage{tablefootnote}

\makeatletter
\let\IEEEorig@makecaption\@makecaption

\renewcommand{\@makecaption}[2]{%
  \ifx\@captype\@IEEEtablestring
    \begingroup
      \let\scshape\relax % disable small caps (your earlier requirement)
      \def\@makecaption##1##2{%
        \vskip\abovecaptionskip
        \setbox\@tempboxa\hbox{##1. ##2}%
        \hspace*{\parindent}{##1. ##2}\par
        \vskip\belowcaptionskip
      }%
      \IEEEorig@makecaption{#1}{#2}%
    \endgroup
  \else
    \IEEEorig@makecaption{#1}{#2}%
  \fi
}
\makeatother

\begin{document}

\title{I can't recognize (yet): Delayed Rendering to Defeat Visual Phishing Detectors}
% What They Can't See:

\input{authors}

\maketitle
\input{sections/0-abstract}

\pagestyle{plain}


\input{structure}

\bibliographystyle{IEEEtran}

{\footnotesize
\input{main.bbl}

%\bibliography{bibliography} 
}

\appendices
\input{appendix/structure_app}

\end{document}

%% file: commands.tex
\newcolumntype{?}{!{\vrule width 1.5pt}}

\newcommand{\textbox}[1]{
    \noindent\fbox{%
        \parbox{0.97\columnwidth}{%
            {#1}
        }%
    }
}

\newcommand\purple[1]{%
  \bgroup
  \hskip0pt\color{purple!50!blue}%
  #1%
  \egroup
}

\newtcolorbox{cooltextbox}[1][]{%
    colback=black!5,
    colframe=black!5,
    notitle,
    sharp corners,
    borderline west={0pt}{0pt}{red!80!black},
    enhanced,
    breakable,
    left=0pt,
    right=0pt,
    top=0pt,
    bottom=0pt
    }

\newcommand\smamath[1]{{\small $#1$}}

\newcommand{\blackcircled}[1]{%
  \tikz[baseline=(c.base)]{
    \node[circle, fill=black, text=white,
          inner sep=0.05ex, 
          minimum size=0.1em 
         ] (c) {\small #1};
  }%
}

%% file: authors.tex
\author{
\hspace{-0mm}
\IEEEauthorblockN{
Ying Yuan\IEEEauthorrefmark{1}, 
Cristiano Alex Rado\IEEEauthorrefmark{2}, 
Giovanni Apruzzese\IEEEauthorrefmark{3}\IEEEauthorrefmark{4}, 
Mauro Conti\IEEEauthorrefmark{2}\IEEEauthorrefmark{5}, 
Luigi Vincenzo Mancini\IEEEauthorrefmark{1}
\\}
\IEEEauthorblockA{\small{ 
\IEEEauthorrefmark{1}\textit{Sapienza University of Rome},
\IEEEauthorrefmark{2}\textit{University of Padua},
\IEEEauthorrefmark{3}\textit{Reykjavik University},
\IEEEauthorrefmark{4}\textit{University of Liechtenstein}, 
\IEEEauthorrefmark{5}\textit{Örebro University}
}}

{
ying.yuan@uniroma1.it, 
cristianoalex.rado@studenti.unipd.it, 
giovannia@ru.is,
mauro.conti@unipd.it,
mancini@di.uniroma1.it
}
}

%% file: sections/0-abstract.tex
\begin{abstract}
Phishing webpages are continuously polluting the Web. By impersonating real-and-trusted services, such malicious websites can deceive users, leading to credential theft or malware download. To cope with such a threat, plenty of countermeasures have been proposed and the most advanced techniques leverage machine-learning methods that infer whether a webpage is benign or not by inspecting its visual representation. The rationale is that a webpage that is (i) visually similar to that of a trustworthy website, but which is (ii) hosted on a domain different from that of such a website, is a clear sign of a phishing webpage. Yet, despite the demonstrated effectiveness of such detection methods, this class of defenses is, by design, susceptible to a specific kind of subtle-but-cheap timing-based attacks which---worryingly, and perhaps surprisingly---have never been investigated so far. Such an oversight questions the overall reliability of these defenses in the wild, and demands a dedicated treatment.

First, we show that timing-based evasion attacks have not been accounted for by prior work on visual phishing website detectors. Then, we elucidate the intrinsic vulnerability of these detectors: they can be bypassed by \textit{delaying the rendering} of webpage elements. Practically, these detectors must compute a visual similarity score between a target webpage and a known legitimate one. This requires taking a ``snapshot'' of the target webpage before the similarity computation.
Attackers can deliberately delay the rendering of key elements, such as the logo, so that these elements appear fully only after the snapshot has been taken. 
This simple tactic misleads the visual-similarity module, leading the system to incorrectly classify the phishing page as benign.
We empirically show that state-of-the-art detectors can be completely defeated (detection rate dropping from 100\% to 0\%) by employing easy-to-apply ``problem-space'' techniques such as curtain effects. We also carry out a user study, evaluating the effectiveness of these attacks against real humans, and find that end users are unable to reliably identify our ``perturbations'' ($p<.05$). Finally, we propose mitigations, including a browser-extension that, without making any call to remote services, warns users that they may have landed on a phishing webpage. We release all of our resources.
\end{abstract}

%% file: structure.tex
\input{sections/1-introduction}
\input{sections/2-related}

\input{sections/3-threat}
\input{sections/4-implementation}
\input{sections/5-results}
\input{sections/6-userstudy}
\input{sections/7-Countermeasures}

\input{sections/8-discussion}

\input{sections/9-conclusions}
\input{sections/11-acknowledgment}

%% file: sections/1-introduction.tex
\section{Introduction}
\label{sec:introduction}

\noindent
Phishing is a prominent threat in the current cybercrime landscape~\cite{fbi2024icr,proofpoint2024phish,apwg2024,barracuda2024report}. In particular, despite decades of research~\cite{dhamija2006phishing, alsharnouby2015phishing, baki2022sixteen}, phishing \textit{websites} continue to pose a serious threat to internet users~\cite{aljofey2022effective,abbasi2021phishing,acharya2025pirates}. To fight back, researchers have proposed numerous blacklist-based~\cite{bell2020analysis,oest2020phishtime} or data-driven anti-phishing schemes~\cite{liu2022inferring, spacephish2022,tian2018needle,rao2020catchphish} that leverage features extracted from URLs (e.g., URL length), HTML structures (e.g., external links) or visual elements (e.g., logo similarity). Defenses conforming to the latter group, thanks to the substantial recent advances in machine learning (ML), received particular attention and are now being leveraged by operational security systems~\cite{apruzzese2023real,draganovic2023users}.

Visual phishing detectors, such as VisualPhishNet~\cite{abdelnabi2020visualphishnet}, PhishPedia~\cite{lin2021phishpedia}, and PhishIntention~\cite{liu2022inferring} have proven to be highly effective in countering phishing websites deployed in the wild. These systems rely on the intuition that attackers craft their phishing websites so that they closely resemble webpages of well-known brands---but which are deployed on domains that are clearly different from the legitimate brand’s domain.
For instance, attackers may want to steal the login credentials of PayPal users by crafting a malicious webpage that is visually similar to that of PayPal, but which is hosted on a domain that is not associated to those of PayPal. Practically, the detection follows a three-step approach. When a user lands on a webpage, {\small \textit{(1)}}~the detector captures a screenshot. {\small \textit{(2)}}~It then computes the visual similarity between key webpage elements, such as the logo, and those included in a predefined ``reference list'' of trusted websites. Then, {\small \textit{(3a)}}~if the webpage is deemed to resemble that of a well-known brand, the system checks whether the webpage is hosted under the legitimate domain of such a brand, flagging the webpage as malicious if the domain does not match, and as benign otherwise. However, {\small \textit{(3b)}}~if no match is found within the reference list, the system considers the webpage as legitimate: this is done to avoid raising false positives, given the fast-paced nature of the Web (e.g., well-known brands update their webpages frequently). Despite some misclassifications, these detectors are appreciated for their efficiency, as they enable to counter the majority of phishing scams~\cite{lin2021phishpedia,liu2022inferring} without making multiple (expensive) queries to third-party services; besides, a recent work even found that visual detectors are even robust to perturbations targeting HTML-based detectors~\cite{hao2024doesn}.

\begin{figure*}[!t]
\centering
\includegraphics[width=\linewidth]{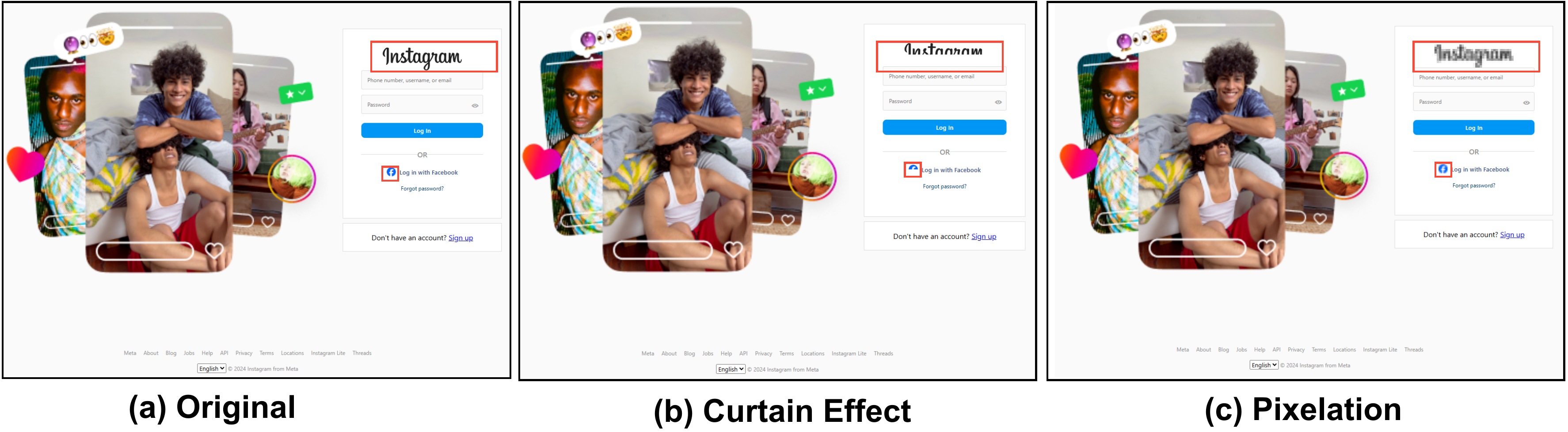}
\vspace{-7mm}
\caption{
\textbf{Examples of timing-based attacks.} We show the original real-world phishing webpage (in Fig.~\ref{fig:adv_ex}(a)) and two adversarial variants produced by delaying the rendering of its logos (see the red boxes): (Fig.~\ref{fig:adv_ex}(b)) curtain effect---the logos progressively renders from top to bottom, (the image shows only its upper half); (Fig.~\ref{fig:adv_ex}(c)) pixelation---logos progressively loads from blurred to clear, (the image displays a blurred version). 
}
\vspace{-3mm}
\label{fig:adv_ex}
\end{figure*} 

Indeed, various prior works investigated the ``adversarial robustness'' of (visual) phishing detectors. Such assessments involved two types of strategies: {\small \textit{(a)}} applying visible manipulations, such as altering logo colors or fonts~\cite{hao2024doesn,ji2025evaluating}; or {\small \textit{(b)}} introducing human-imperceptible perturbations through well-known adversarial ML techniques~\cite{kurakin2016adversarial, lee2023attacking}. In either case, the overarching principle was the same: bypassing the component tasked to compute the visual similarity between the given webpage and those in the reference list. Crucially, however, all such prior work assumed that \textit{webpages render correctly and screenshots are accurately captured}. Such an assumption overlooks the dynamic nature of webpage loading, where UI elements such as logos and background images require time to appear. 
For reference, the screenshot-capture component of PhishIntention~\cite{liu2022inferring} is set to trigger after 2s, but a concurrent study found that, on average, webpages load in 7.2s~\cite{csontos2021accessibility}. \textit{Disruptions during this loading period can result in incomplete or incorrect screenshots}, thereby reducing the effectiveness of the visual-similarity component of the phishing detector. Worryingly, such disruptions can also be deliberately introduced by attackers, who may craft phishing webpages so that UI elements load only after a deliberate delay---potentially in a fashion that would not make users more suspicious (e.g., by progressively improving the quality of the element).

\vspace{1mm}
\noindent
\textsc{\textbf{Summary and Research Questions.}} We begin our study by asking a motivational research question (RQ0) to substantiate our scientific novelty: ``\textit{\purple{To what extent has prior work accounted for `timing attacks' against visual phishing detectors?}}''. We address RQ0 via a systematic literature review, and find that prior work overlooked the intrinsic security vulnerability of visual phishing detectors to timing attacks and that, more broadly, scarce consideration was given to potential disruptions occurring before the screenshot is captured. 

Based on this finding, we posit RQ1: ``\textit{\purple{What are some ways in which delayed-rendering techniques can be exploited to bypass visual phishing detectors?}}'' To this end, we propose a generic threat model for timing-based attacks and hypothesize two practical implementation strategies: {\small \textit{(a)}}~curtain effects, which seek to load webpage elements progressively from top to bottom; and {\small \textit{(b)}}~pixelation techniques, which seek to progressively render a given image by increasing the number of pixels displayed.   Fig.~\ref{fig:adv_ex} shows examples of such rendering manipulations.

Next, and as our major technical contribution, we consider RQ2: ``\textit{\purple{What impact do such delayed rendering techniques have against state-of-the-art visual detectors?}}'' We hence take four state-of-the-art (and publicly-available) visual phishing detectors~\cite{abdelnabi2020visualphishnet,lin2021phishpedia,liu2022inferring,liu2024less} and test them, in an end-to-end fashion, against our proposed attacks. First, we collect a dataset of 2,000 recent phishing webpages from the APWG eCrime Exchange~\cite{apwgecx}, and identify 991 malicious webpages suitable for our study. Then, to derive a baseline, we submit these webpages to PhishIntention~\cite{liu2022inferring}, flagging those correctly identified as phishing. Next, we apply our envisioned delayed-rendering techniques, and craft a total of 880 adversarial webpages---each conforming to different variants of our envisioned attacks (e.g., slower or faster rendering, more/less pronounced pixelation). Finally, we submit them to our four considered detectors, and measure the impact: in some instances, such as when the logo is barely rendered (25\% visibility), the detection rate of PhishIntention~\cite{liu2022inferring} and Phishpedia~\cite{lin2021phishpedia} drops to 0\%, and even PhishLLM~\cite{liu2024less} only achieves a 30\% detection rate; whereas VisualPhishNet~\cite{abdelnabi2020visualphishnet} (which looks at the entire screenshot, and not just to the logo) appears to be robust, but we found that this is due to an artifact that can be trivially circumvented.

Such findings prompted us to investigate the impact of our attacks against humans (RQ3): ``\textit{\purple{Can humans spot such `rendering-based perturbations' and, if so, when?}}'' Through a user study (n=247), we assessed {\small \textit{(i)}}~if and when humans can perceive that our delayed-rendering techniques lead to noticeable differences in a given webpage---in a non-primed setup; and {\small \textit{(ii)}}~whether the user would be suspicious of a webpage with a, potentially, altered rendering process. We found that, generally, users notice that something is different when elements take more than 3s to load fully, and that 70\% of our adversarial webpages were trusted by our participants.

To mitigate the problem we brought to light, we wonder (RQ4): ``\textit{\purple{What are some possible countermeasures to such a threat?}}'' We argue that a practical defense can be provided by a browser extension which checks if a given webpage has some elements which are designed to render in an ``adversarial'' way. We hence develop a prototype of such a browser extension, which operates locally---without making any remote call to third-party services. To our knowledge, we are the first to propose a similar plugin to cover a blind spot of visual phishing detectors.

\vspace{1mm}
\noindent
\textbf{\textsc{Contributions.}} 
After substantiating the research gap tackled by this paper (in Section §\ref{sec:related}):
\begin{itemize}[leftmargin=*]
    \item We elucidate the vulnerability of visual phishing detectors to \textit{delayed-rendering} attacks~(§\ref{sec:threat}). We are the first to assess the end-to-end impact of similar attacks against four state-of-the-art systems~(§\ref{sec:implementation}).

    \item Our empirical evaluation (§\ref{sec:results}) confirms that existing systems are \textit{intrinsically vulnerable} to such a threat: we found statistically-significant (\smamath{p\!<\!.05}) drops in detection rates. To shed more light on this issue, we investigate the human perspective of delayed-rendering attacks (§\ref{sec:userstudy}).
    
    \item We propose and develop a browser-extension plugin to mitigate this problem (§\ref{sec:countermeasure}). Technically, we release our resources (dataset and source code) in~\cite{repository}.
   
\end{itemize}
Our findings should motivate more research to focus on this subtle issue (more discussions are in §\ref{sec:discussion}).

%% file: sections/2-related.tex
\section{Motivation and Related Work}
\label{sec:related} 
\noindent
Phishing attacks use deceptive ``lures" to trick victims into carrying out tasks favorable to cybercriminals, such as revealing sensitive information (e.g., credit card numbers, or login credentials) or infecting systems with malware~\cite{IBMphish2025,deloitte2025,stivala2023attachments}. By leveraging social-engineering techniques, attackers impersonate legitimate entities, conveying fraudulent content through emails (spear phishing)~\cite{lee2021d,evans2022raider,thapa2023evaluation}, SMS messages (smishing)~\cite{sanjari2024smishviz}, phone calls (vishing)~\cite{figueiredo2025sounds,sharevski2025blind}, QR codes (quishing)~\cite{weinz2025impact}, or malicious websites~\cite{spacephish2022}. According to IBM~\cite{IBM2025}, phishing accounts for 15\% of breaches and each incident costs \smamath{\approx}\$5M to organizations.

In this work, we focus on phishing websites. We first summarize this specific threat~(§\ref{ssec:addressing}), then motivate our paper via a systematic literature review (§\ref{ssec:slr}), and finally define the problem tackled by our research (§\ref{ssec:gap}).

\subsection{Addressing Phishing Websites}
\label{ssec:addressing}
\noindent
Phishing websites are a never-ending problem in the Web ecosystem~\cite{cic2024phishing}. This is because phishing websites can be deployed at scale and at no risk by attackers~\cite{spacephish2022}. 

There exist numerous ways to mitigate the spread of phishing websites. Alongside manually adding known malicious URLs to popular blocklists (which can be leveraged by Web browsers~\cite{gsb} or organization-wide firewalls~\cite{weinz2025impact}), data-driven defenses include analysing the URL~\cite{maneriker2021urltran} or the HTML~\cite{lim2024phishing} of a webpage. However, such mechanisms are costly to deploy: for instance, the hybrid detector of SpacePhish~\cite{spacephish2022} requires making multiple queries to third-party services (e.g., WHOIS and DNS queries, retrieving the pagerank, as well as inquiring various reputational-based services). To address such a shortcoming, recent works proposed anti-phishing schemes relying on the visual representation of webpages, and which require only minimal resources to operate. 
Indeed, while the idea of detecing phishing webpages via visual-similarity-based methods is over 20 years old (e.g.,~\cite{wenyin2005detection}), operational deployment of similar techniques became possible only in the 2020s, thanks to recent advances in ML and computing~\cite{draganovic2023users, divakaran2022phishing,lin2021phishpedia}. 

Modern visual phishing detectors leverage the fact that, to trick users, most phishing websites are crafted to resemble reputable services, despite being deployed on different domains (e.g., a phishing webpage resembling that of PayPal can be deployed on a domain such as www.paypa1.org). State-of-the-art detectors (such as VisualPhishNet~\cite{abdelnabi2020visualphishnet}, Phishpedia~\cite{lin2021phishpedia}, PhishIntention~\cite{liu2022inferring}, or PhishLLM~\cite{liu2024less}) seek to counter phishing webpages mimicing those within a shortlist of popular brands that are typically targeted by phishers (e.g., Paypal, or Facebook). Specifically, these detectors use ML models to quickly (\smamath{<}1s~\cite{liu2022inferring}) compare the visual similarity of various elements of a given webpage, such as the logo, to those contained in a reference list based on the protected brands (e.g.,  for PayPal, the reference list can have multiple logos). If a match is found,\footnote{A ``match'' occurs if the visual similarity between the two elements is above a threshold according to a given visual-similarity metric.} then this could be indicative of either {\small \textit{(a)}}~a phishing webpage trying to imitate a protected brand, or {\small \textit{(b)}}~a legitimate webpage of the protected brand: to verify this, a single ``domain check'' is carried out, leading to either a malicious (if the webpage is not hosted on a domain of the ``matched'' brand) or benign (otherwise) decision. However, if there is no match between the elements of the given webpage and any of those in the reference list, then such detectors would treat the webpage as benign: this is to avoid raising false positives in the case of, e.g., a given brand updates its logo, or a webpage of a brand outside those in the shortlist~\cite{lee2023attacking}.

Visual phishing detectors are being used by security companies~\cite{draganovic2023users, apruzzese2023real}. Yet, like any ML-based method, they are prone to ``adversarial examples''~\cite{spacephish2022}. Therefore, various studies (e.g.,~\cite{ji2025evaluating, roh2025evaluating}) have investigated the robustness of such detectors to adversarial perturbations. For instance, the creators of Phishpedia~\cite{lin2021phishpedia} and PhishIntention~\cite{liu2022inferring} evaluated their systems against well-known gradient-based strategies; whereas Hao et al.~\cite{hao2024doesn} used diffusion models to generate semantically-similar logos, and Lee et al.~\cite{lee2023attacking} proposed generative adversarial perturbations that deceived even human users. These, and others (e.g.,~\cite{ji2025evaluating}) evasion attempts can be mitigated by well-known adversarial ML defenses such as adversarial training~\cite{hao2024doesn}.

\subsection{Systematic Literature Review (RQ0)}
\label{ssec:slr}

\noindent
Security evaluations of visual phishing detectors always focused on evading the visual-similarity model---e.g., by introducing manipulations that lead the model to output a similarity below the threshold needed to obtain a ``match'' with an element inside the reference list. However, as we argued, this class of phishing detectors can be bypassed via delayed-rendering techniques, which fall outside the adversarial ML domain. Here, we provide factual evidence that such strategies have been overlooked by prior work.

\textbf{Approach.} To investigate RQ0, we conduct a systematic literature review (SLR). First, we identified papers that either {\small \textit{(a)}}~proposed or {\small \textit{(b)}}~assessed the security of visual phishing detectors. To this end, we considered five recent seminal works~\cite{abdelnabi2020visualphishnet, lin2021phishpedia, liu2022inferring, liu2024less, ji2025evaluating} and applied the snowball method~\cite{wohlin2014guidelines} (as also done, e.g., in a recent SoK~\cite{schroer2025sok}); we complemented this search by looking for papers mentioning ``phishing'' on DBLP. We specifically focused on peer-reviewed (excluding, e.g., unpublished works) and technical (excluding, e.g., literature reviews) papers that specifically focused on the aforementioned topics. Determining the topic fit was done by two authors who qualitatively inspected the content of the paper, and shared their findings to reach a consensus. For each work that fit our scope, we carried out full-text keyword searches~\cite{radhakrishnan2017novel, weinz2025impact}, looking for occurrences of the following terms: {\small \{``time,'' ``capture,'' ``timing,'' ``timeout,'' ``render''\}}. Whenever we found a match, we reviewed the context of each occurrence to determine whether the paper discussed \textit{how} or \textit{when} visual elements were `acquired' before being processed by the envisioned detection system. In addition, we further examine the methodology section of each paper to infer how screenshot was captured.

\begin{table}[t]
\caption{{\textbf{Literature Overview.} Papers on ``visual phishing detection”  found via our SLR. For each work, we report: whether it considers that the rendering process can be exploited, whether the method to capture the screenshot is explained, and the capturing time (if provided).}} 
\label{tab:literature} 
\vspace{-2mm}
\small
    \centering
    \resizebox{\linewidth}{!}{%
        \begin{tabular}{c|c|c|c|c}
            \toprule
            \begin{tabular}{c}\textbf{Paper} \\ \textbf{(1st Author)}\end{tabular} & \textbf{Year} & \begin{tabular}{c}\textbf{Keywords} \\ \textbf{Relevant}\end{tabular} & \begin{tabular}{c}\textbf{Screenshot} \\ \textbf{acquisition}\end{tabular} & \begin{tabular}{c}\textbf{Capture} \\ \textbf{time}\end{tabular} \\
            \midrule
            Medvet~\cite{medvet2008visual} & 2008 & \ding{51} & \ding{51} &-\\
            Liang~\cite{liang2016cracking}&2016& \ding{51} &  \ding{51} &-\\
            Corona~\cite{Deltaphish}&2017&  & \ding{51} &-\\
            Abdelnabi~\cite{abdelnabi2020visualphishnet}&2020&  & &-\\ 
            % \midrule
            
            Lin~\cite{lin2021phishpedia}&2021&  & \ding{51} &1.88s\\
            % \midrule
             Liu~\cite{liu2022inferring}&2022&  & \ding{51} &2s\\
             % \midrule
             Lee~\cite{lee2023attacking}&2023&  & &-\\
             % \midrule
             Miao~\cite{miao2023good}&2023& \ding{51} &  \ding{51} &-\\
             % \midrule
             Liu~\cite{liu2023knowledge}&2023&  & \ding{51} &-\\
             % \midrul
             Van Dooremal~\cite{van2021combining}&2021& \ding{51} &  \ding{51} &-\\
             % \midrule202
             
             % \midrule
             Jain~\cite{jain2017phishing}&2017&  & &-\\
             % \midrule
             Al-Ahmadi~\cite{al2020deep}&2020&  & \ding{51} &-\\
             % \midrule
             Rao~\cite{rao2020two}&2020&  & &-\\
             % \midrule
             Pandey~\cite{pandey2023phish}&2023&  & \ding{51} &2.2s\\
             % \midrule
             Bozkir~\cite{bozkir2020logosense}&2020&  & &-\\
             % \midrule
             Li~\cite{li2020webpage}&2020&  & &-\\
             % \midrule

             Afroz~\cite{afroz2011phishzoo}&2011& & &-\\
            
             % \midrule
             Panda~\cite{panda2022novel}&2022&  & &-\\
             % \midrule
             
             % \midrule
             Liu~\cite{liu2024less}&2024&  & &-\\
             % \midrule
             Pourmohamad~\cite{pourmohamad2024deep}&2024& \ding{51} &  \ding{51} & -\\
             % \midrule
             Liu~\cite{li2024knowphish}&2024&  & &-\\
             % \midrule
             Hao~\cite{hao2024doesn}&2024&  & &-\\
             Roh~\cite{roh2025evaluating}&2025& \ding{51} &  \ding{51} &-\\
             Ji~\cite{ji2025evaluating}&2025& \ding{51} &  & -\\
            \bottomrule
        \end{tabular} 
        }

\vspace{-6mm}
\end{table}

\textbf{Findings (overview).} We report the results in Table~\ref{tab:literature}, showing the 24 most relevant works we found. Specifically, for each work, we report whether our considered keywords were deemed ``relevant'' for RQ0 (e.g., an ``irrelevant'' occurrence would be if ``time'' was used to denote ``training time''); whether the paper provided details on how the screenshot was captured (which could indicate a potential acknowledgment of the susceptibility to rendering issues); and whether the paper specified ``when'' the screenshot was captured (we even looked at the source-code, if available). While 12 (out of 24) papers do explain the screenshot-capturing process, the majority of the works (20 out of 24) do not provide low-level information on how long the considered detector ``waits'' before capturing the screenshot: only three papers~\cite{lin2021phishpedia,liu2022inferring,pandey2023phish} specify the capturing time of the screenshot, which \smamath{\approx}2s, i.e., well-below the average loading time of webpages (7.2s, as measured in~\cite{csontos2021accessibility}). This already indicates that such methods may be prone to naturally-occurring rendering issues. We even reached out to the authors of~\cite{draganovic2023users}, who confirmed that the partnered company does indeed face similar problems in the Wild. 

\textbf{Low-level analysis.} Let us discuss our qualitative analyses on the contextual relevance of the keyword occurrences. First, we could not find relevant matches for 16 (out of 24) papers: for instance, Liu et al.~\cite{liu2022inferring} mention ``time'' three times, ``render'' once, ``capture'' 8 times, and ``timeout'' once---but do so in contexts that do not presume the presence of attackers (e.g., ``Generation of phishing websites is largely automated and causes a time delay between blacklist updates and zero-day phishing emergence''); similarly,~\cite{liu2023knowledge} state ``We observed that some websites take longer  to respond to our visit, exceeding our predefined timeout. [...] We suggest that practitioners can adjust the timeout to a longer duration'' which does not imply the possibility of deliberate attacks. Among the 7 papers for which we could find matches relevant for RQ0, we report:~\cite{van2021combining}, stating ``By relying on a screenshot of the [rendered] web page (rather than relying only on features extracted from the web page), our approach is robust against resource evasion techniques, such as image splitting'' which, however, does not account for delayed-rendering attacks; ~\cite{medvet2008visual}, stating ``some differences may be voluntarily inserted, both at the source level or at the rendering level'', which acknowledges the potential for rendering-based attacks (despite not carrying out any evaluation); and~\cite{pourmohamad2024deep}, stating ``The extraction of DOM and Term features is constrained by a 500-millisecond timeout. If this duration is exceeded, classification is aborted, potentially allowing exploitation by attackers'', which also acknowledge such a possibility despite not investigating it further (note: this capture is \textit{not related} to the screenshot, explaining why we did not report it in Table~\ref{tab:literature}). In summary, even though some works do mention that attackers may bypass detectors by exploiting the rendering process, no prior work practically evaluated the impact of such a threat.

\begin{cooltextbox}
    \textbf{\textsc{Answer to RQ0:}} Prior work proposing, or evaluating the security of, visual phishing detectors assumed that the screenshot used to compute the visual similarity is taken when all elements in a given webpage are fully rendered. Even though some works acknowledged that this assumption can be exploited by adversaries, we could not find practical evaluations of this vulnerability.
\end{cooltextbox}

\subsection{Research Gap and Problem Statement}
\label{ssec:gap}
\noindent
Our SLR highlights a critical gap in the security of visual phishing detectors: to the best of our knowledge, the impact of timing-based attacks against such detectors is still unknown. We seek to address this open problem.

First, we propose a generic threat model for this particular class of evasion attacks. Next, we practically evaluate the effectiveness of these attacks against state-of-the-art visual phishing detectors. Then, and as recommended by~\cite{yuan2024adversarial, draganovic2023users}, we will also assess if adversarial webpages that bypassed our considered detectors can also deceive human users. Finally, we will provide a client-side countermeasure to mitigate this vulnerability.

\vspace{1mm}
\textbox{{\textbf{Disclaimer:} The ways in which rendering-based attacks can be physically realized is infinite, and cannot be evaluated in a single publication. This is why our evaluation encompasses two exemplary ways of carrying out such attacks, for which we operate in the problem space~\cite{Pierazzi2020IntriguingSpace}.}}

%% file: sections/3-threat.tex
\section{Threat Model}
\label{sec:threat}
\noindent
We define our threat model for delayed-rendering attacks. Following the recommendation of~\cite{apruzzese2023real}, we describe our envisioned scenario by providing the system-wide perspective (instead of assuming an ML-centric view).

\subsection{Target System}
\label{ssec:system}
\noindent
We consider the setting in which a defender seeks to detect phishing webpages via visual phishing detectors---encompassing either logo-based (e.g., PhishIntention~\cite{liu2022inferring}) and full-page-based (e.g., VisualPhishNet~\cite{abdelnabi2020visualphishnet}) systems. These systems require the definition of a \textit{reference list} containing visual elements (i.e., logos or full-page screenshot) referring to brands that are typically targeted by phishers: the tacit assumption is that phishing websites targeting brands outside the reference list are uncommon and hence are less problematic (since fewer users can fall victim to them). For instance, PhishIntention's reference list spans 277 brands and \smamath{\approx}4k logos~\cite{liu2022inferring}. 

\begin{figure}[t]
\centering
\includegraphics[width=\linewidth]{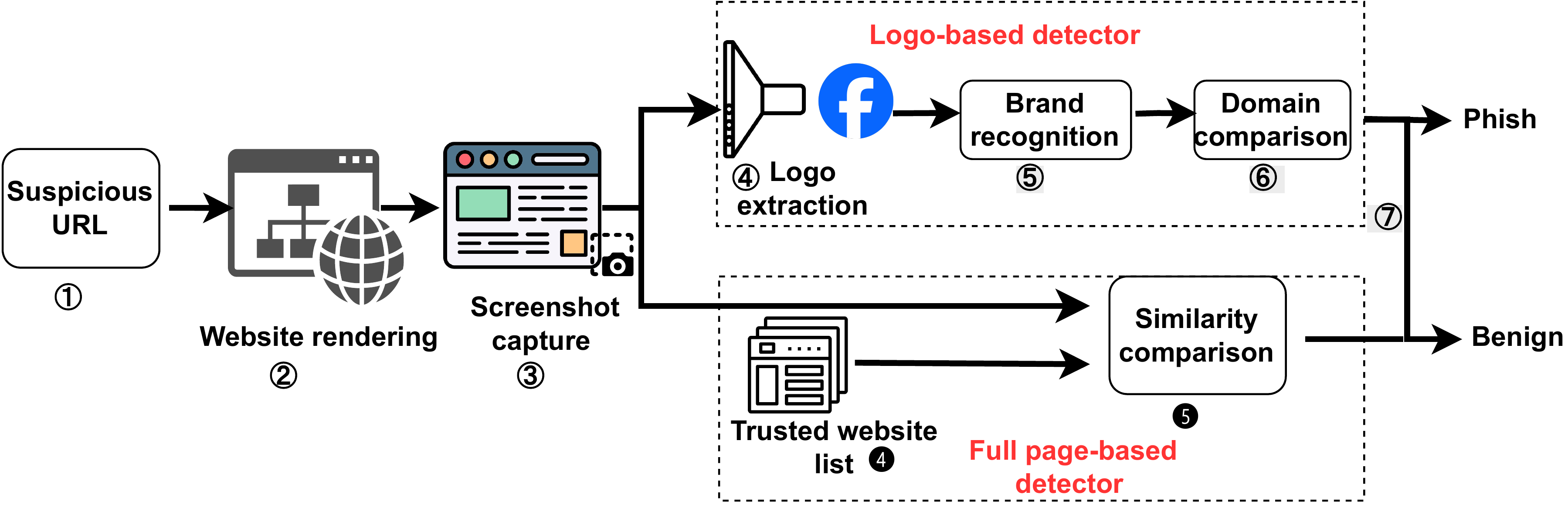}
\vspace{-0.2in}
\caption{
\textbf{Visual Phishing Detector}---%
   {We show the workflow of logo-based and full page-based visual phishing detectors.} 
}
\vspace{-0.2in}
\label{fig:visual_detector_workflow}
\end{figure}

As shown in Fig.~\ref{fig:visual_detector_workflow}, visual phishing detectors follow a common workflow: upon receiving (\ding{172}) a ``suspicious'' URL (i.e., an URL for which it is unsure whether it is benign or malicious), the corresponding website is automatically visited (\ding{173}), and a screenshot is captured for the subsequent similarity analysis (\ding{174}).

Logo-based detectors employ ML–based object recognition techniques (e.g., Siamese neural networks and OCR) to identify and extract logos (\ding{175}) from screenshots. Then, vision-based approaches are used to recognize if the logo(s) pertain to one of the target brand (\ding{176}). This can be done in many ways. For instance, PhishIntention~\cite{liu2022inferring} first compares the cosine similarity between the extracted logo(s) and each logo in the reference list, and the logo having the highest similarity (provided that it is above the threshold of \smamath{\theta=0.87}) is considered as that of the brand impersonated by the webpage; at the same time, OCR can also be used to extract textual elements and augment the brand-identification process. Other detectors (e.g.,~\cite{liu2024less}) may also leverage LLM for this task. Then, after identifying the target brand, a DNS query (\ding{177}) is issued to check if the webpage is hosted under a domain of the targeted brand. If the two differ, the page is classified as phishing; otherwise, it is deemed benign (\ding{178}).

Full page-based detectors analyze the visual similarity of entire webpage screenshots. Instead of using a reference list containing ``logos of brands'', these detectors rely on a ``trusted website list'' (\blackcircled{4}), containing screenshots of webpages known to be benign. Hence, after receiving the screenshot of a suspicious webpage, these detectors calculates its similarity (e.g., via Siamese neural nets for VisualPhishNet~\cite{abdelnabi2020visualphishnet}) with those in the trusted website list (\blackcircled{5}). If the minimum distance between the suspicious webpage and websites in the trusted list smaller than a predefined threshold (determined during the models training stage), the webpage is classified as phishing, indicating an attempt to impersonate one of the trusted websites with high visual similarity. Otherwise, if the distance is not small enough, the page is classified as benign.

\subsection{Envisioned Attacker}
\label{ssec:attacker}
\noindent
We define our attacker with well-known notation~\cite{biggio2018wild}.

\textbf{Goal.} The attacker wants to evade visual phishing detectors, i.e., inducing test-time misclassifications wherein phishing webpages are deemed as benign. The attacker wants to craft phishing webpages mimicking popular brands---i.e., resembling those of the brands included in the detectors reference list. Finally, the attacker wants to ensure that the adversarial webpages are still perceived as legitimate by humans.

\textbf{Knowledge.} The attacker knows of the existence of visual phishing detectors ``protecting'' the brands that the attacker wants to impersonate in the phishing webpages. The attacker is also aware of the state-of-the-art, and hence knows that such systems must eventually capture a screenshot of a suspicious webpage. The attacker expects such a screenshot to be taken after 2s (based on Table~\ref{tab:literature}). The attacker, however, does not know anything about the specific internal functionalities of the detection systems, and cannot even access the classification output (according to the terminology in~\cite{apruzzese2023real}, this is an ``invisible'' ML system from the attacker's viewpoint).

\textbf{Capabilities.} The attacker has no access to the targeted system. The attacker cannot make queries, manipulate the training data, or modify the reference list. The attacker has complete control over the phishing webpages they can create---which represent the ``problem space'' (see~\cite{spacephish2022, Pierazzi2020IntriguingSpace}). Finally, the attacker can deploy the webpage on URLs not yet included in a blocklist (otherwise, the attack would have no effect whatsoever~\cite{weinz2025impact}).

\subsection{Adversarial Strategies [RQ1]}
\label{ssec:strategy}
\noindent
Within the aforementioned scenario, the attacker exploits the intrinsic vulnerability of visual-phishing detectors to delayed-rendering evasion attempts. Indeed, given the scarce consideration that prior work has given to such a threat (see §\ref{ssec:slr}), and given that other forms of adversarial ML attacks can be countered via ad-hoc countermeasures (see §\ref{ssec:addressing}), this is a sensible strategy.

As we stated (in §\ref{ssec:gap}) there are many ways in which an attacker can deliberately tamper with the rendering process of their controlled phishing webpages. For instance, an attacker may introduce overlays; or add banners above certain elements (e.g., logos), potentially making them automatically disappear (even gradually) after a short timer---all with the purpose of inducing the visual detector to capture a screenshot that does not represent the ``actual intent'' of the phishing webpage. 
In this context, we envision two noteworthy strategies that attackers may leverage to convey such ``adversarial perturbations'':
\begin{itemize}[leftmargin=*]
    \item \textbf{Curtain effect.} The attacker delays webpage rendering, making UI elements (e.g., logos and background images) to appear progressively in a top-to-bottom, curtain-style effect. As a result, the screenshots provide an incomplete representation of the webpage, e.g., showing only the top 0\%, 25\%, 50\%, or 75\% of a logo or background image (as shown in Fig.~\ref{fig:adv_ex}b). Incomplete logos disrupt the logo extraction (\ding{175}) process, inducing mismatches in the brand recognition pipeline (\ding{176}), due to similarity scores below the threshold, while OCR fails to extract accurate textual descriptions, leading LLM-based inference to misidentify the brand (\ding{177}). Similarly, partially rendered background images increase the embedding distance from legitimate full-page screenshots in the trusted list (\blackcircled{4}), pushing similarity scores above the threshold used by full-page detectors (\blackcircled{5}). Consequently, adversarial screenshots generated through curtain-effect rendering delays compromise both logo-based and full-page visual detectors, enabling evasion.

    \item \textbf{Pixelation.} The attacker leverages a lazy-loading strategy to make logos and background images gradually transition from blurred to clear, intentionally slowing webpage rendering. As a result, web crawlers capture screenshots (\ding{174}) containing pixelated logos and backgrounds (e.g., with block sizes of 5×5, 4×4, 3×3, or 2×2 pixels, as shown in Fig.~\ref{fig:adv_ex}c). The expected effects of such screenshots to the visual phishing detector are the same as those of the aforementioned ``curtain effect''. 
\end{itemize}
Both of these techniques involve problem-space perturbations that are physically realizable and within the capabilities of our attacker, since they can be implemented via simple javascript (as we will show in §\ref{sec:implementation}). Finally, note that these techniques can even be combined (e.g., progressively showing a pixelated version of a logo).

\begin{cooltextbox}
    \textbf{\textsc{Answer to RQ1:}} There are many ways to exploit delayed-rendering techniques to evade visual phishing detectors. We posit that ``curtain effects'' and ``pixelation'' are two practical strategies. This is because not only they can have (in theory) an impact on the detectors' performance, but also because they can be overlooked by humans---who, even if they notice something, may believe that the delayed rendering is simply a naturally-occurring phenomenon.
\end{cooltextbox}

%% file: sections/4-implementation.tex
\section{Implementation}
\label{sec:implementation}

\noindent 
To assess the impact of our threat model, we first describe the data-collection procedure (§\ref{ssec:data}), which we use to re-create four state-of-the-art visual phishing detectors (§\ref{ssec:detectors}), and to provide a basis for implementing our envisioned delayed-rendering strategies (§\ref{ssec:attack}).

\subsection{Data Collection}
\label{ssec:data}

\noindent
Prior work proposed various datasets of phishing websites (e.g.,~\cite{liu2022inferring,lin2021phishpedia,liu2024less,li2024knowphish}). However, these (and others, such as~\cite{ji2025evaluating}) datasets are not appropriate for our assessment. 

First, because---to implement our attacks---we need data providing, for each webpage, its HTML, JavaScript, CSS, and all UI components (e.g., icons and logos). Unfortunately, the datasets used in most related works~\cite{liu2022inferring,lin2021phishpedia,liu2024less,li2024knowphish,liu2023knowledge} include only URLs, HTML, and screenshots, lacking the JavaScript and UI components required to reproduce webpage rendering offline (i.e., without downloading external elements, which have been taken down, leading to elements which fail to load). This problem also affects more recent datasets~\cite{ji2025evaluating,lee2024beneath}. Other datasets (e.g., those used in~\cite{lee2024beneath, pourmohamad2024deep}) are outdated, and therefore do not represent valid samples for our user study. 

Therefore, we constructed a new phishing dataset sourced from the Anti-Phishing Working Group and, specifically, from its eCrime Exchange~\cite{apwgecx} platform (as also done by prior works~\cite{ji2025evaluating,lee2024beneath,li2024knowphish}). After receiving permission, we implemented a crawler that queried APWG eCX every 30 minutes to retrieve live phishing URLs. 
Between May 5--14, 2025, our crawler collected 2,000 raw URLs with APWG confidence scores $\ge 90$. 
For each URL we captured screenshots with WebDriver~\cite{webdriver} and simulated human interactions to save the complete webpage (including HTML, JavaScript, CSS, and all UI assets) with PyAutoGUI~\cite{pyautogui}, enabling offline analysis. 

Then, we performed data-cleaning operations in three stages: {\small \textit{(1)}}~we inspected saved HTML files and removed obvious error pages, such as pages including ``404'' and ``page not found''; {\small \textit{(2)}}~to empty and duplicate pages, screenshots were clustered by visual similarity using Fastdup~\cite{Fastdup}, and we manually removed redundant data; finally, {\small \textit{(3)}}~we conducted a final manual audit to confirm that all webpages met our requirements. After cleaning, the dataset comprised 991 phishing webpages, covering 162 brands (of which 48 overlap with PhishIntention~\cite{liu2022inferring}).

\subsection{System Creation}
\label{ssec:detectors}

\noindent
To provide a comprehensive (and realistic) evaluation, we consider state-of-the-art visual phishing detectors mimicing those used by real-world companies~\cite{draganovic2023users,apruzzese2023real}.

To this end, and as a byproduct of our literature review (§\ref{ssec:slr}), we considered recent works that proposed visual phishing detectors validated on real webpages and with publicly-available implementations. For instance, we excluded KnowPhish~\cite{li2024knowphish} and Deltaphish~\cite{Deltaphish} because they are not publicly available, and SpacePhish~\cite{spacephish2022} because it does not focus on visual elements. We eventually set for four visual phishing detection systems, published in top-tier venues: VisualPhishNet~\cite{abdelnabi2020visualphishnet}, PhishPedia~\cite{lin2021phishpedia}, PhishIntention~\cite{liu2022inferring}, and PhishLLM~\cite{liu2024less}. Importantly, we use the exact versions provided by the creators.

\begin{figure*}[t]
\centering
\includegraphics[width=\linewidth]{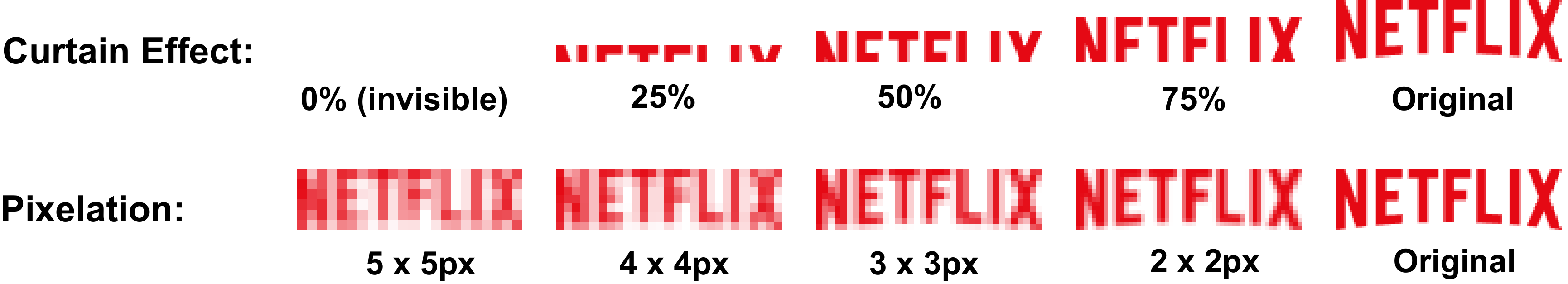}
\vspace{-4mm}
\caption{
\textbf{Adversarial Logo Images.} During webpage rendering delays, we apply curtain-effect and pixelation attacks to the logo images. (a) \textit{Curtain Effect}: logos gradually appear from top to bottom at 0\% (invisible), 25\%, 50\%, 75\%, and 100\% visibility. (b) \textit{Pixelation}: logos transition from blurred to clear as the pixel block size decreases from 5 × 5px to 4 × 4px, 3 × 3px, 2 × 2px, and finally the original resolution.}
\label{fig:adv_logo_ex}
\vspace{-4mm}
\end{figure*} 

For a preliminary evaluation, we use PhishIntention as the main ``baseline'' detector due to its widespread use in the research community\footnote{As of April 2024, the paper~\cite{liu2022inferring} has over 130 citations on Google Scholar since its publication in USENIX SEC'22} but also because it resembles real-world systems~\cite{draganovic2023users, apruzzese2023real}, its source-code is publicly available, and was subject of various robustness assessments (e.g.,~\cite{hao2024doesn,lee2023attacking, ji2025evaluating}) but against different forms of adversarial strategies. We hence submit the 991 phishing samples in our dataset to PhishIntention, and find that 125 are correctly classified as phishing (the remaining are either of brands outside the reference list, or too recent). So, we will use these 125 phishing webpages as the basis for crafting adversarial phishing samples.\footnote{Note: our goal is assessing if our considered strategies leads to evasion of pages that would otherwise be detected. In the Appendix~\ref{app:extend}, and as a sanity check, we also test if webpages that are \textit{not flagged} as phishing \textit{still evade} the detection after applying our modifications.}

To quantitatively measure our attacks, we define two metrics: the False-Negative Rate ($FNR$), representing the proportion of malicious samples that are incorrectly classified as benign; and the Attack-Success Rate ($ASR$) which is the difference between the $FNR$ computed on the ``adversarially-manipulated'' webpages and their original variants (i.e., $ASR=FNR(atk)-FNR(no\_atk)$). We do not need to measure false-positive rates: we do not focus on availability attacks, and given that we use the systems provided by (reputable and peer-reviewed) prior work, we can trust that these systems work well.

\subsection{Attack Realization}
\label{ssec:attack}

\noindent
In what follows, we explain how we practically realize our ``adversarially-manipulated'' webpages by applying curtain and pixelation effects to their original variants. 

First, we ensure each webpage deploys correctly offline with all resources intact and no broken images. We do so manually, fixing loading issues by downloading missing images from the original online website or correcting resource paths, and store logos and background images in a local components folder to enable reliable offline rendering on modern browsers; this additional verification step led to excluding 101 phishing samples due to rendering errors in the original webpage (which we did not notice during our initial cleaning operations). Then, in the remaining 24 webpages -- which represent those used as a basis for our evaluation -- we insert unique identifier `\textit{id}' attributes into `\verb|<div>|' and `\verb|<img>|' tags within the HTML to programmatically control the loading of these UI elements. Doing so enabled us to develop `\textbf{PhishMe}', a JavaScript module (provided in our repository) that automatically deploys our delayed-rendering attacks by manipulating the HTML files of our considered webpages.

Our implementation of the \textbf{curtain effect} seeks to gradually reveal logos or background images from top to bottom while delaying webpage rendering. To achieve this, \textit{PhishMe} first retrieves each image's dimensions (width and height) and then applies the CSS property `\verb|clip-path: inset(0px, 0px, value)|' to conceal a specified percentage of the image's bottom portion while preserving its layout. The script incrementally decreases the hidden bottom area over time (e.g., revealing an additional 25\% of the height per second), thereby progressively increasing the visible top region. We divide the rendering process into five stages--hiding 100\%, 75\%, 50\%, 25\%, and 0\% of the bottom area, respectively, which transitions the image from completely invisible to fully visible, as shown in the first row of Fig.~\ref{fig:adv_logo_ex}.

Websites often display blurred placeholder images that are subsequently replaced with higher-resolution versions to improve perceived loading speed. Our \textbf{Pixelation} attack leverages this method by pixelating logo and background images while intentionally delaying webpage rendering, thereby generating adversarial screenshots with degraded visual quality. To quantitatively regulate the manipulation, we define uniform transformations that reduce image resolution to 1/2, 1/3, 1/4, and 1/5 of the original resolution by two steps: \textit{downscale} and \textit{upscale}. Specifically, \textit{PhishMe} first retrieves image’s width and height, then downscales the images by a factor of \textit{1/N} (i.e., 1/2, 1/3, 1/4, 1/5) using bilinear interpolation, which computes each new pixel’s color as a distance-weighted average of the four nearest pixels in the original image, producing smooth and blurred transitions. Subsequently, these downscaled images are upscaled back to their original dimensions using nearest-neighbor resampling, which duplicates each pixel from the downscaled images into larger \textit{N\smamath{\times}N} (i.e., 2\smamath{\times}2px, 3\smamath{\times}3px, 4\smamath{\times}4px, 5\smamath{\times}5px) pixel blocks, resulting in pixelated effect. \textit{PhishMe} progressively loads these pixelated logos and background images, transitioning from heavily blurred to clear, to generate adversarial variants. An example of our pixelation can be seen in the second row of Fig.~\ref{fig:adv_logo_ex}, where the logo images gradually sharpen from blurred to clear.

\vspace*{0.1in}
\textbox{\textbf{Remark.} The gradual shift (e.g., going from 2\smamath{\times}2 to 3\smamath{\times}3px) can be implemented quickly or slowly. In our evaluation, we will consider that the screenshot is captured at each ``interval'' (so to cover all possible cases).}

%% file: sections/5-results.tex
\section{Impact Assessment [RQ2]}
\label{sec:results}
\noindent
We quantitatively assess the impact of our attacks against state-of-the-art detectors. First, we preliminary test if our hypotheses are, in principle, correct~(§\ref{ssec:pre_test}) and then proceed with the main evaluation~(§\ref{ssec:main_test}).

\subsection{Screenshot manipulation vs PhishIntention}
\label{ssec:pre_test}

\noindent 
\textit{Do our attacks work---in principle?} Before committing to a large experimental campaign, we examine if PhishIntention can be evaded by a proof-of-concept implementation of our attacks---done by applying the manipulations at the screenshot level. (This assessment is easy to carry out, since we only need to change a single image to simulate the effect of a delayed-rendering technique.)

\textbf{Method.} We take the 125 webpage screenshots that are classified as phishing by PhishIntention (§\ref{ssec:data}) and apply our visual perturbations to the entire screenshot (e.g., we pixelate everything), and then submit such screenshots to PhishIntention. Even though such changes are conspicuous, they are doable and we are merely examining if they lead to any difference in the detection performance. We hence create, for each of these 125 screenshots, a total of 1,000 variants (because we consider four different ``intensities'' for both the pixelation and curtain effects).

\begin{figure}[t]
\centering
\includegraphics[width=\linewidth]{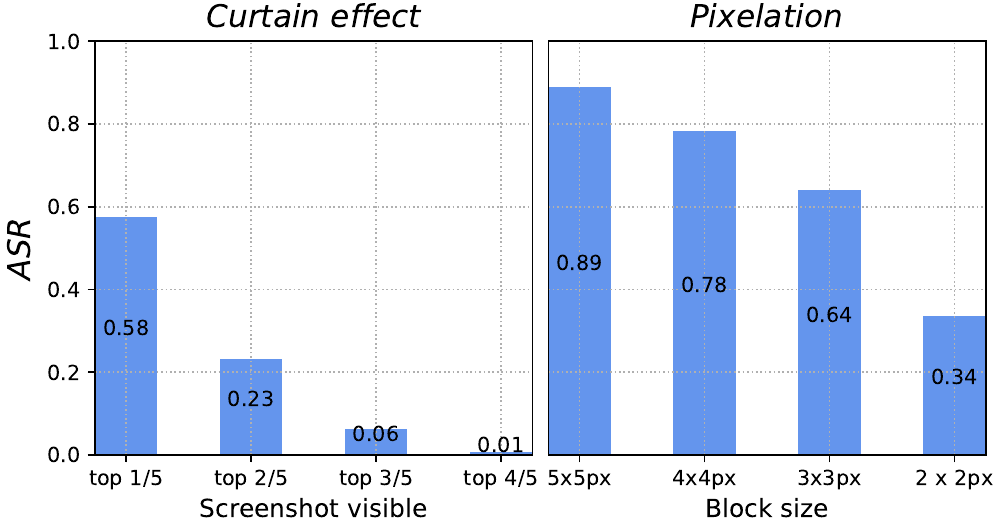}
\vspace{-7mm}
\caption{
\textbf{Attack Effectiveness on Screenshot.} We directly apply \textit{Curtain effect} and \textit{Pixelation} attacks to screenshot images. For the Curtain effect, we display the top 1/5, 2/5, 3/5, and 4/5 of each screenshot. For Pixelation, we pixelate screenshots with 5×5, 4×4, 3×3, and 2×2 pixel blocks. We then report the attack success rate (\textit{ASR}) against PhishIntention. 
}
\vspace{-0.2in}
\label{fig:pre_result}
\end{figure}

\textbf{Results.} We report the results in Fig.~\ref{fig:pre_result}, showing the $ASR$ achieved by each type (and intensity) of adversarial strategy. Recall that all these 125 webpages identified by PhishIntention (i.e., baseline $ASR=0$). We see that the $ASR$ increases linearly with the intensity. Under the \textit{Curtain effect}, as the visible portion of the screenshot decreased from 4/5 to 1/5, the $ASR$ increased from 0.01 to 0.58. Similarly, under \textit{Pixelation} attack, as screenshots became more blurred (from pixelated with 2\smamath{\times}2 to 5\smamath{\times}5px blocks), the $ASR$ increased linearly from 0.34 to 0.89. To statistically assess attacks effectiveness, we conducted pairwise Chi-squared tests to compare the number of misclassified adversarial phishing samples with the number of misclassified original phishing samples. The differences were statistically significant (\smamath{p<.01} for all pairs) except for screenshots with 4/5 visibility generated by \textit{Curtain effect}, confirming the effectiveness of our attacks.

\vspace*{0.1in}
\textbox{\textbf{Takeaway.} Our preliminary evaluation indicates that PhishIntention is likely susceptible to our attacks.}

\subsection{Main Evaluation: On Website}
\label{ssec:main_test}
\noindent 
For the main evaluation, we simulated a real-world setup. 

\textbf{Method.} We deployed all 24 webpages listed in Table~\ref{tab:dataset} on our local server (to avoid polluting the Web), and used \textit{PhishMe} to apply our attacks during the webpages' rendering process. 
We applied \textit{Curtain effect} and \textit{Pixelation} attacks to the logo images of all webpages, and to background images when they existed (i.e., 10 webpages). Our preliminary evaluation demonstrates that screenshots with 4/5 visibility generated by \textit{Curtain effect} attack did not noticeably affect detector performance. In the main study, we captured screenshots slightly earlier to create adversarial screenshots with 0\%, 25\%, 50\%, and 75\% visibility (instead of 20\%, 40\%, 60\% and 80\%) of both logo and background images. \textit{Pixelation} attacks remain the same as in the preliminary study (i.e., pixelated blocks of 2×2px, 3×3px, 4×4px and 5×5px) and were applied to both logos and backgrounds. 

\textbf{Expectation.} We implemented a total of 60 attack variants: 4 \textit{Curtain effect} (§\ref{sssec:curtain}), 4 \textit{Pixelation} (§\ref{sssec:pixelation}), and 12 combination attacks (\textit{Curtain effect} + \textit{Pixelation}), applied to logos, backgrounds, or both simultaneously (§\ref{sssec:comb}). All variants were evaluated against four visual phishing detectors. We hypothesize that logo-based detectors (e.g., PhishPedia~\cite{lin2021phishpedia} or PhishIntention~\cite{liu2022inferring}) would be more (less) impacted by changes to the logo (background) than full-page detectors (VisualPhishNet~\cite{abdelnabi2020visualphishnet}).

\subsubsection{Curtain Effect impact}
\label{sssec:curtain}
\begin{figure}[t]
\centering
\includegraphics[width=\linewidth]{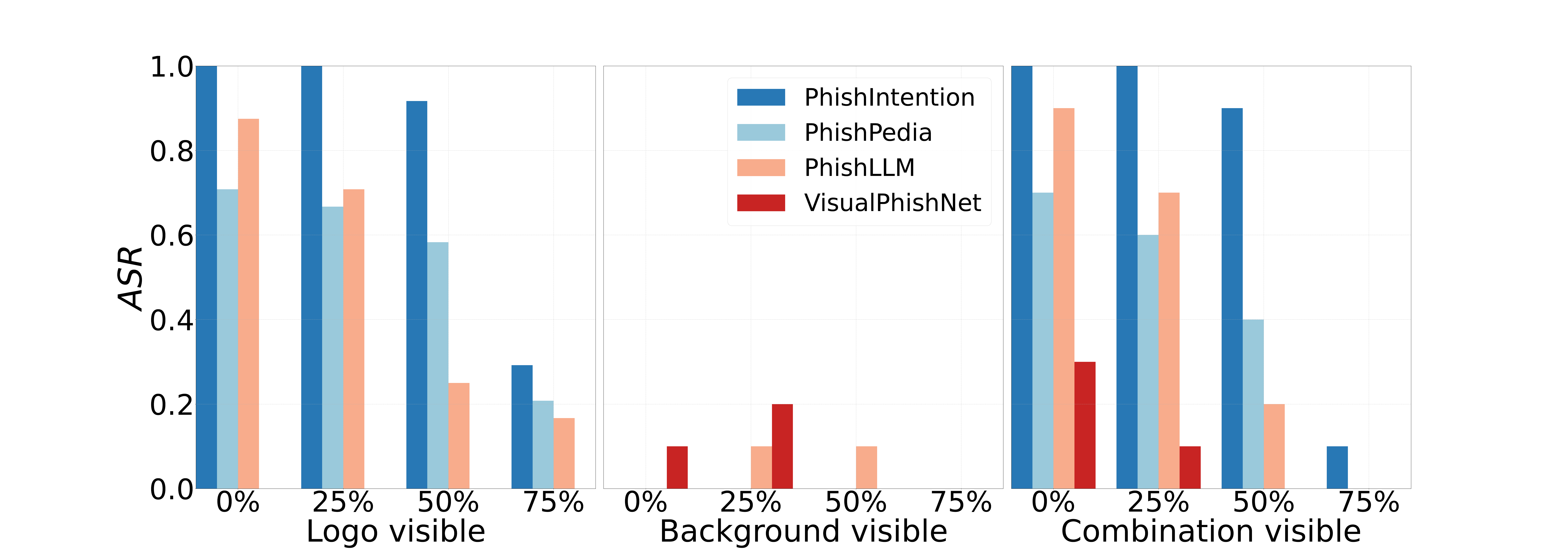}
\vspace{-5mm}
\caption{
\textbf{Curtain effect Effectiveness (on Website).} We report the attack success rate of four visual phishing detectors against adversarial screenshots, that are captured during the rendering process of logo, background, and both images, with visibility ranging from 0\% to 75\%.}
\vspace{-4mm}
\label{fig:curtain_result}
\end{figure}

Our results demonstrate that logo-based visual phishing detectors are highly vulnerable to \textit{Curtain Effect} attacks. As shown in Fig.~\ref{fig:curtain_result} (Logo visible), when screenshots were captured with logos at 0\% or 25\% visibility, all 24 phishing samples successfully bypassed both PhishIntention and PhishPedia (\smamath{ASR=100\%}). Even PhishLLM, the most recent visual detector based on LLM, was bypassed (\smamath{ASR\approx70\%}) at these visibility levels. Moreover, the attack effectiveness shown an approximately linear decline as logo visibility increases in these logo-based detectors.

In contrast, attacking logos does not degrade the performance of full page-based visual phishing detector: VisualPhishNet. Instead, some adversarial samples become easier to be detected than the original phishing samples. For example, the false negative rate (FNR) decreases from 0.375 to 0.292 (refer to Table~\ref{tab:curtain_asr}) when the webpage logo is rendered invisible under the \textit{Curtain Effect}. To understand why this happens, we examined VisualPhishNet's workflow. The model compares each screenshot against a target list and uses the visual distance to determine whether a page is phishing. Because logos are key cues for target identification, hiding or altering them could cause the model to match the page to a different target with a more similar layout. This incorrect match reduces the visual distance and moves it below the model's decision threshold, leading to a correct phishing classification \footnote{As shown in Fig.~\ref{fig:visual_samples}, when the Outlook logo becomes invisible, the model switches its match from RBC Bank to PayPal, whose layout is more similar, yielding a smaller visual distance that falls below the threshold and triggers a correct phishing prediction.}.

However, when applying \textit{Curtain effect} attacks on background images, the \smamath{ASR} on VisualPhishNet reaches 20\% (refer to `Background visible' in Fig.~\ref{fig:curtain_result}). 
Background manipulation (e.g., displaying only 25\% of background images) reduces similarity with potential targets, causing misclassifications. Background manipulation does not affect logo similarity-based detectors like PhishIntention and PhishPedia, but decreases PhishLLM's performance by up to \smamath{10\%} due to its reliance on broader visual context.

\vspace*{0.1in}
\textbox{\textbf{Takeaway.} Our \textit{Curtain effect} attacks on logos significantly degrade the performance of logo-based detectors, with \smamath{ASR} decreasing in a roughly linear manner as logo rendering. This reveals that the most effective attack window occurs during the initial stage of logo rendering.}

\begin{figure}[t]
\centering
\includegraphics[width=\linewidth]{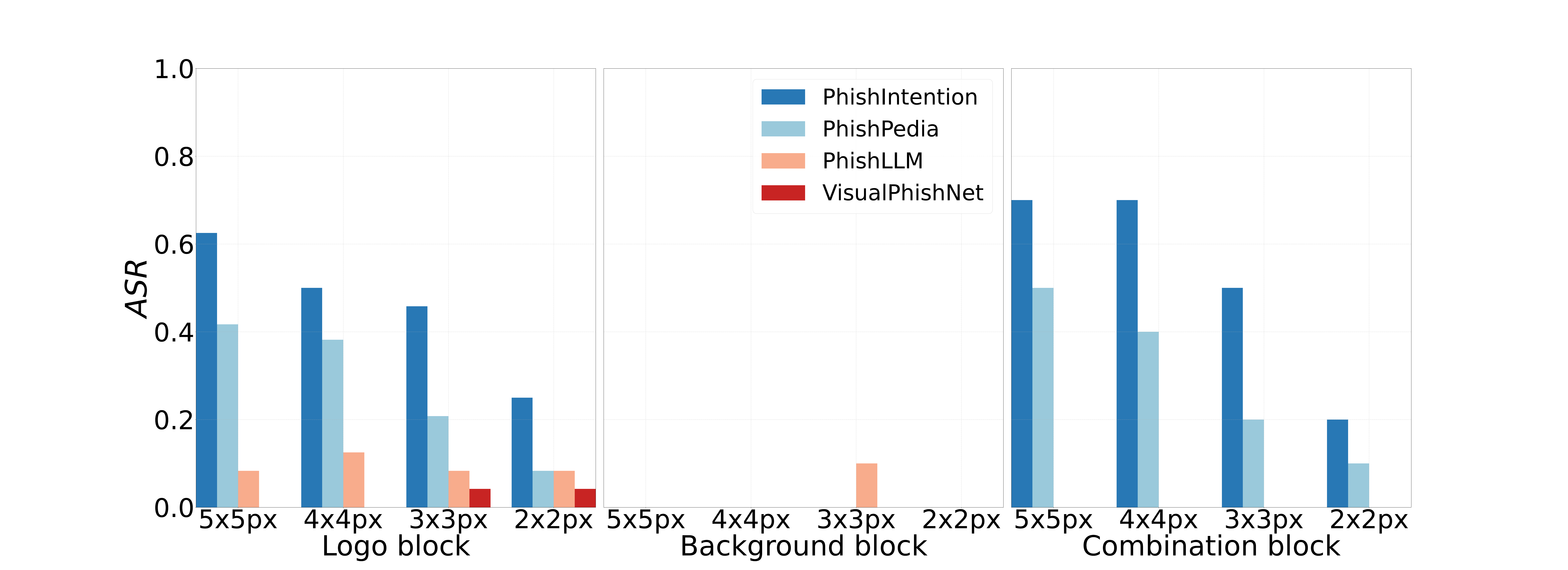}
\vspace{-0.2in}
\caption{
\textbf{Pixelation Effectiveness (on Website).} We report the attack success rates of four visual phishing detectors against adversarial samples generated by \textit{Pixelation} attacks. These adversarial screenshots are captured during the rendering of logo images, background images, or both, with pixelation block sizes ranging from 2×2 to 5×5 pixels.} 
\vspace{-0.2in}
\label{fig:pixelation_result}
\end{figure} 

\subsubsection{Pixelation impact}
\label{sssec:pixelation}
Our results in Fig.~\ref{fig:pixelation_result} show that logo similarity-based detectors are highly vulnerable to \textit{Pixelation} attacks when logos (or both logos and backgrounds) are pixelated. The highest \smamath{ASR}, 0.62 for PhishIntention and 0.50 for PhishPedia, occur when logos are pixelated with 5\smamath{\times}5px blocks, producing the strongest blurring effect. The overall trend is consistent with that of \textit{Curtain effect} attacks: the \smamath{ASR} decreases approximately linearly as attack intensity decreases from 5\smamath{\times}5 to 2\smamath{\times}2 pixels (i.e., as images transition from blurred to clear).

However, \textit{Pixelation} attacks are less effective than \textit{Curtain effect} attacks (which achieve 100\% \smamath{ASR}). This is due to insufficient pixelation on large images. Specifically, logos have different resolutions, so applying the same pixelation blocks will create different levels of blur. High-resolution logos remain sufficiently clear even after pixelation, preventing the attack from achieving the same degree of effectiveness at deceiving the detector. %complete blurring effect similar to the invisibility created by \textit{Curtain effect} attacks.

When pixelating background images, the attacks are effective only against PhishLLM, reaching a 10\% \smamath{ASR} with 3\smamath{\times}3px blocks. This confirms that the pixelation strength is insufficient to meaningfully distort large images such as backgrounds.
When pixelating both the logos and backgrounds simultaneously, the attacks are effective only against PhishIntention and PhishPedia, producing slightly higher success rates than pixelating the logos alone. 

\vspace{0.1in}
\textbox{\textbf{Takeaway.}
Pixelation attacks are more effective when applied to logos (w.r.t. background) images, with their effectiveness influenced by image size and resolution. However, according to our results, pixelation is statistically significantly (\smamath{p<.05}) weaker than curtain effects.
}

\subsubsection{Pixelation+Curtain Effect impact}
\label{sssec:comb}
We implemented 36 combination attacks that simultaneously apply \textit{Curtain effect} and \textit{Pixelation} with varying attack intensities (e.g., 5\smamath{\times}5 pixels \& 25\% visibility) to logos, backgrounds, or both. We report in Table~\ref{tab:combine_asr} (in the Appendix) the \smamath{FNR} to measure the percentage of adversarial webpages that successfully bypassed detection.

In general, such combination attacks achieved up to 100\% evasion against PhishIntention, PhishPedia, and PhishLLM, while 50\% of samples bypassed the full page-based detector VisualPhishNet. These results demonstrate the effectiveness of our attacks and highlight the serious security risks posed by this overlooked vulnerability in webpage rendering: practically implementing these ``combined'' attacks is trivial by expert attackers.

\begin{cooltextbox}
    \textbf{\textsc{Answer to RQ2:}} Simple, low-cost attacks entailing (combinations of) pixelation and curtain effects applied during webpage rendering can bypass state-of-the-art visual phishing detectors systems. Logo-based detectors are particularly prone (\smamath{ASR\rightarrow100\%}), whereas detectors ingesting the whole-page screenshot are more robust (\smamath{ASR\rightarrow20\%}) but are also less effective in non-adversarial scenarios~\cite{liu2022inferring}.
\end{cooltextbox}

%% file: sections/6-userstudy.tex
\section{User Study: are humans fooled, too?}
\label{sec:userstudy}
\noindent
After demonstrating that delayed webpage rendering attacks can fool visual phishing detectors, we conduct a user study to investigate whether these attacks also deceive human users---the true target of phishing. 
Indeed, phishing attacks should simultaneously~\cite{yuan2024adversarial} deceive both human users and automatic detection systems---which is also one of our envisioned attacker's goals.\footnote{We observe that few works (e.g.,~\cite{lee2023attacking,hao2024doesn}) carry out a security evaluation that also encompasses a user study. For instance, recent works accepted in top-tier venues (e.g.,~\cite{ji2025evaluating, roh2025evaluating}) do not do so.}

We provide an overview (§\ref{ssec:survey_design}), discuss technical details and recruitment (§\ref{ssec:survey_details}) and present the results (§\ref{ssec:survey_result}).

\subsection{Design and Overview}
\label{ssec:survey_design}
\noindent 
Let us outline the key elements of our user study.

\textbf{Objective and Idea.} At a high level, our user study seeks to answer RQ3, which can be disentangled in RQ3a ``\textit{\purple{Can `unaware' humans notice our considered delayed-rendering attacks?}}'' and RQ3b ``\textit{\purple{Are human more suspicious of webpages that appear to load more slowly?}}''. The underlying principle is that, in the real-world, webpage rendering does not occur instantaneously and internet users may overlook that certain elements load more slowly and, even if they do notice, they may not suspect that it is a way to conceal a phishing attack. Nevertheless, to simulate a real-world environment, we record videos that show the common process of {\small \textit{(i)}}~clicking on a given click-through button, and {\small \textit{(ii)}}~show the entire rendering phase of one of our considered webpages. Indeed, users do not see static screenshots or isolated UI elements (such as logos or background images) when they interact with the Web.

\textbf{Design novelty.} Our user study differs from those carried out by prior work. Specifically, we both {\small \textit{(i)}}~investigate if users can spot some changes in the rendering phase---without priming them about phishing, and hence representing a realistic scenario; and {\small \textit{(ii)}}~examine if users aware of potential phishing tactics are suspicious or not of delayed-rendering elements in a webpage---representing a ``worst-case'' scenario for an attacker. We are not aware of prior works that carried out these two distinct types of investigations within the same paper (e.g.,~\cite{hao2024doesn} used the latter approach, whereas~\cite{lee2023attacking} followed the former).

\textbf{Structure.} We devised a questionnaire consisting in four segments (I--IV) designed to answer RQ3a and RQ3b. 
\begin{enumerate}[label=\Roman*), leftmargin=4mm]
    \item Participants are shown a consent form that provides general information about the study, including the procedures, participants' tasks, and rights. To avoid priming, we do not mention terms such as ``phishing'' or ``suspicious'' or ``trust'' at this stage. After consenting, participants watch a test video showing flowing water and are asked whether it plays smoothly on their device. This step ensures that our following videos render properly and prevent biases caused by poor internet connection, which could affect participants' perception.\footnote{If participants \textit{always} have a poor connection, then they would be naturally prone to our attacks---but this should not be the common case.} If playback is not smooth, the survey ends.

    \item Participants watch five pairs of videos, each corresponding to a different website/brand. Each pair consists of one video showing the webpage rendering at ``normal speed'' (we used PingDom~\cite{Pingdom} to infer that the average loading time for popular websites visited from the US West Coast is 1s) and another video showing the same webpage with rendering delayed by a randomly assigned duration (2, 3, 4, or 5 seconds). For each pair, we ask participants: \textit{``How would you rate the similarity of the website's loading process shown in these two videos?''} (similarly to~\cite{lee2023attacking}) using a 4-point Likert scale: 1 (Not Similar), 2 (Somewhat Similar), 3 (Similar), and 4 (Identical), which does not include a neutral option to encourage participants make a clear judgment; altogether these questions serve to answer RQ3a. Next, participants are asked \textit{``What is the name of the brand shown in the video?"} which is answerable in a textbox. This question, besides being useful for the ecological validity of our study, also serves as an attention check to identify low-quality responses from participants who may not be paying close attention. To avoid forcing guesses for unfamiliar brands, we explicitly instruct: \textit{``If you cannot recognize the brand, write `don't know'".} We also ask a follow-up question \textit{``How familiar are you with this brand?"} (similarly~to~\cite{hao2024doesn}).

    \item Here, we inform participant that the following tasks envision a ``phishing'' scenario. Specifically, we inform participants that phishing websites \textit{``may try to tamper with the loading process to evade detectors"}, which places participants in a highly-primed setting to estimate the upper-bound performance of users (used in~\cite{hao2024doesn,yuan2024adversarial}). Then, participants watch 25 videos covering 5 website brands, with 5 videos per brand. Each set of five videos uses the same rendering time but different attack types, randomly selected from \textit{36} variants (refer to §\ref{ssec:survey_details}) (e.g., Netflix with a pixelated logo rendering in 3 seconds). We randomly combine brand names, attack variants, and rendering times to ensure each participant sees a unique set of videos, such that each one watches the same videos only once. For consistency, we ensure that each participant watches the same number of attack types. After each video, a participants must answer: \textit{``If you were visiting the webpage of this well-known website, how much would you trust such a webpage if its rendering process resembled what was shown in the video?"}, using a 4-point Likert scale: 1 (Distrust), 2 (Somewhat Distrust), 3 (Somewhat Trust), and 4 (Trust). After every five videos, participants were asked to report the brand name and their familiarity with it (as done in the previous segment).

    \item Lastly, participants answer demographic questions, including gender, age-range, education level, and whether they have a background in computer science or cybersecurity (as also done in~\cite{hao2024doesn}).

\end{enumerate}
Overall, each participant views 35 videos covering 10 website brands. The five brands shown for Q3a are websites without background images, where the videos display only logo-based rendering attacks, as these performed best in automatic detection systems (based on our prior evaluation in §\ref{sec:results}). The other five brands for Q3b include background images, showing attacks applied to both logos and backgrounds. To avoid order bias~\cite{ferber1952order}, the video sequence is randomized for each participant.

\subsection{Details and Recruitment}
\label{ssec:survey_details}
\noindent 
We provide additional details on our design choices.

\textbf{Selection of Phishing Websites.} Recall (from §\ref{sec:implementation}), that our main evaluation spans across 24 websites from 18 brands (shown in Table~\ref{tab:dataset}). Most website brands are from the U.S. To conduct meaningful research, we select well-known U.S. brands based on the popularity rankings from Similarweb~\cite{similarweb} and Semrush~\cite{semrush}. Additionally, given that our user study evaluates attacks on both logos and background images, the candidate websites should include these visual elements (either logos or both logo and background). Accordingly, we select 10 popular brands (five with background images and five without) that are frequently targeted by phishing attacks (according to recent CloudFlare reports~\cite{cloudflare2025top}): Outlook, eBay, Spotify, Instagram (with backgrounds), DHL, Comcast, Facebook, Netflix, PayPal, and Wells Fargo.

\textbf{Videos Creation.} Given the 60 attack variants (§\ref{ssec:main_test}), including \textit{Curtain effect}, \textit{Pixelation}, and combination attacks targeting logos, backgrounds, and both, we recorded videos displaying webpage rendering under these attacks with different rendering times, controled by our script. To simulate realistic webpage rendering, we first identified the legitimate website targets of the 10 phishing websites, and measured their normal rendering times using Pingdom~\cite{Pingdom}, a publicly available platform that has been used in prior work for webpage loading speed measurement~\cite{bakar2024uncovering}. For each legitimate webpage's URL, we ran five tests using Pingdom's San Francisco server and obtained an average rendering time of 0.962 seconds. Based on this, we set 1 second as the normal (baseline) rendering speed in our videos.
For consistency, all videos are \textit{8} seconds long with the same structure: a 2-second warm-up showing a button clicking, a 5-second webpage rendering (1s for normal speed or 2--5s for delayed rendering with attacks), and a 1-second ending. Specifically, for the five websites with background images, we generated 45 videos per brand, covering \textit{Curtain effect}, \textit{Pixelation}, and combination attacks applied to logos, backgrounds, or both\footnote{i.e., 5 (rendering time: 1-5s) * (3 (logo attack types: cutrain effect, pixelation and combination) + 3 (background attack types: curtain effect, pixelation and combination) + 3 (curtain effect, pixelation and combination attacks on both logo and backgrounds))=45}. For the other five websites without background images, we generated 15 videos per brand, consisting of \textit{Curtain effect}, \textit{Pixelation}, and combination attacks applied to logos only\footnote{i.e., 5 (rendering times:1-5s) * 3 (logo attack types: cutrain effect, pixelation and combination)=15}. Overall, we created 300 videos. \footnote{i.e., 5 brands * 45 + 5 brands * 15=300}.

\textbf{Recruitment and Ethics.} Our study was reviewed and approved by our university's research data advisor, who oversees research-related ethical compliance, including GDPR, information security, and research data management. Because our study does not collect personally identifiable information (PII) or sensitive data, a formal Institutional Review Board (IRB) approval is not required at our University. Nonetheless, our study follows the Menlo Report~\cite{bailey2012menlo}: we do not publish phishing or adversarial phishing websites online---only show videos demonstrating the webpage rendering process to participants. We also responsibly informed our participants, which gave us their explicit consent; participants could drop out at any point in time (even after completion) and their responses would not have been recorded.
During our study, we recruited participants from U.S. through Prolific~\cite{prolific}, a widely used crowdsourcing platform, known for providing higher-quality responses than alternatives such as MTurk~\cite{palan2018prolific}. All participants are voluntary and anonymous. Each participant receives \$2.75 (\$11/hour) for completing the survey, meeting Prolific's wage requirements (minimum \$8/hour).
We recruited $n$=250 participants. After removing low-quality responses identified through attention check questions, 247 valid responses remained. We report the demographics of participants in Table~\ref{tab:demographic_accuracy}. On average, participants spent 16.48 minutes completing the study.

\subsection{User Study Results [RQ3]}
\label{ssec:survey_result}
\noindent 
Given our sample size, we can answer RQ3 quantitatively.

\textbf{RQ3b: are users suspicious?}
To answer RQ3b, we primed participants who then watched 25 videos showing webpage rendering under different attacks. In total, we collected responses about users' trust for 6,175 videos. The distribution of responses across attacks and rendering delays is shown in Table~\ref{tab:video_p2} (in the Appendix). We report the overall \textit{User Trust Rate} as the percentage of videos that participants rated as ``trust" or ``somewhat trust" out of all videos shown for RQ3b, which can be seen as a proxy for the \smamath{ASR} of our attacks against human users. The overall user trust rate is 72\%, meaning that \textit{72\% of our adversarial webpages deceived human users} (who considered them as ``trusted''). We ran a \textit{t-test}, confirming that participants obviously trust these adversarial samples, with \smamath{p<.01}. This demonstrates that our attacks were significantly not perceived by human users.

\begin{table}[t]
\caption{\textbf{User Trust Rate Across Rendering Times and Attacks.} We report user's trust rates (combining ``Somewhat Trust" and ``Trust" responses) for videos displaying webpage rendering with different delay times and attack types.}
\label{tab:user_trust} 
\vspace{-3mm}
\small
    \centering
    \resizebox{\linewidth}{!}{%
        \begin{tabular}{c|c|c|c|c|c}
            \toprule
            \textbf{Attack target} &\textbf{Attacks}& \textbf{2s} & \textbf{3s} & \textbf{4s}&\textbf{5s}\\
            \midrule

            \multirow{3}{*}{Logo} & Curtain effect&0.78&0.74&0.74&0.71\\ \cline{2-6}
           & Pixelation& \textbf{0.83}&0.80&0.76&0.77\\ \cline{2-6}
           & Combination &0.78&0.76&0.69&0.66 \\ \hline
           
        \multirow{3}{*}{Background} & Curtain effect&0.77&0.74&0.76&0.71\\ \cline{2-6}
           & Pixelation&0.79&0.80&0.80&0.79 \\ \cline{2-6}
           & Combination &0.79&0.71&0.72&0.69 \\ \hline

        \multirow{3}{*}{Combination} & Curtain effect&0.70&0.75&0.64&0.62\\ \cline{2-6}
           & Pixelation&0.73&0.69&0.69&0.66 \\ \cline{2-6}
           & Combination &0.72&0.63&0.61&\textbf{0.57} \\
            \bottomrule
        \end{tabular} 
        }
        \vspace{-3mm}
\end{table}

To further investigate the effectiveness of different attacks on users, we analyzed users’ trust rates for different adversarial webpages. As shown in Table~\ref{tab:user_trust}, we report users' trust rates for adversarial samples generated by 36 different attack variants. Overall, participants were generally likely to trust the adversarial webpages perturbed by our attacks (with user trust rates around 70\% across all attack variants). The lowest user trust rate is 57\%, observed for the combined attack with a 5-second delayed rendering time--the strongest attack in our study--which is still slightly above random guessing (50\%). The highest attack success rate is \textit{83\%}, for videos showing webpage rendering with pixelation attacks on the logo image and with a 2-second delay. Our findings show that our attacks are consistently effective against both human users and automatic detection systems. For example, logo-based attacks are more likely to bypass automatic phishing detectors, as discussed in §\ref{ssec:main_test} (the \smamath{ASR\rightarrow100\%}). Participants also showed high trust for adversarial webpages having logo attacks, with trust rates reaching up to 83\% for pixelated logo images.

Finally, we recall that RQ3b assume a high-priming setting (§\ref{ssec:survey_design}) in which participants were explicitly prepared to detect phishing websites with delayed rendering. This design represents the least favorable condition for our attacks, probably leading to the `worst performance' of our attacks. In real-world scenarios, where users are not primed to `phishing website with delayed rendering', our attacks would likely be even more effective~\cite{hu2021assessing}.

\begin{cooltextbox}
    \textbf{\textsc{Answer to RQ3}b:} Approximately 70\% of our adversarial webpages were trusted by participants, indicating that our delayed webpage rendering attacks can also significantly evade human detection. The attack effectiveness varied slightly across different attack types.
\end{cooltextbox}

\textbf{RQ3a: can users see it (and when)?} Our results for RQ3b show that most of our adversarial webpages can bypass human detection, but about 30\% still raised user suspicions about the webpages' trustworthiness. To answer RQ3a, we showed each participant 5 pairs of videos: original phishing webpages rendered at normal speed and the webpages rendered with delays of 2, 3, 4, or 5 seconds. Overall, our participants watched 2,470 videos (1,235 pairs), leading to 1,235 responses about whether the rendering of the adversarial webpage was perceived as similar to the normal rendering. The distribution of sample counts is reported in Table~\ref{tab:video_p1} (in the Appendix).

\begin{figure}[t]
\centering
\includegraphics[width=\linewidth]{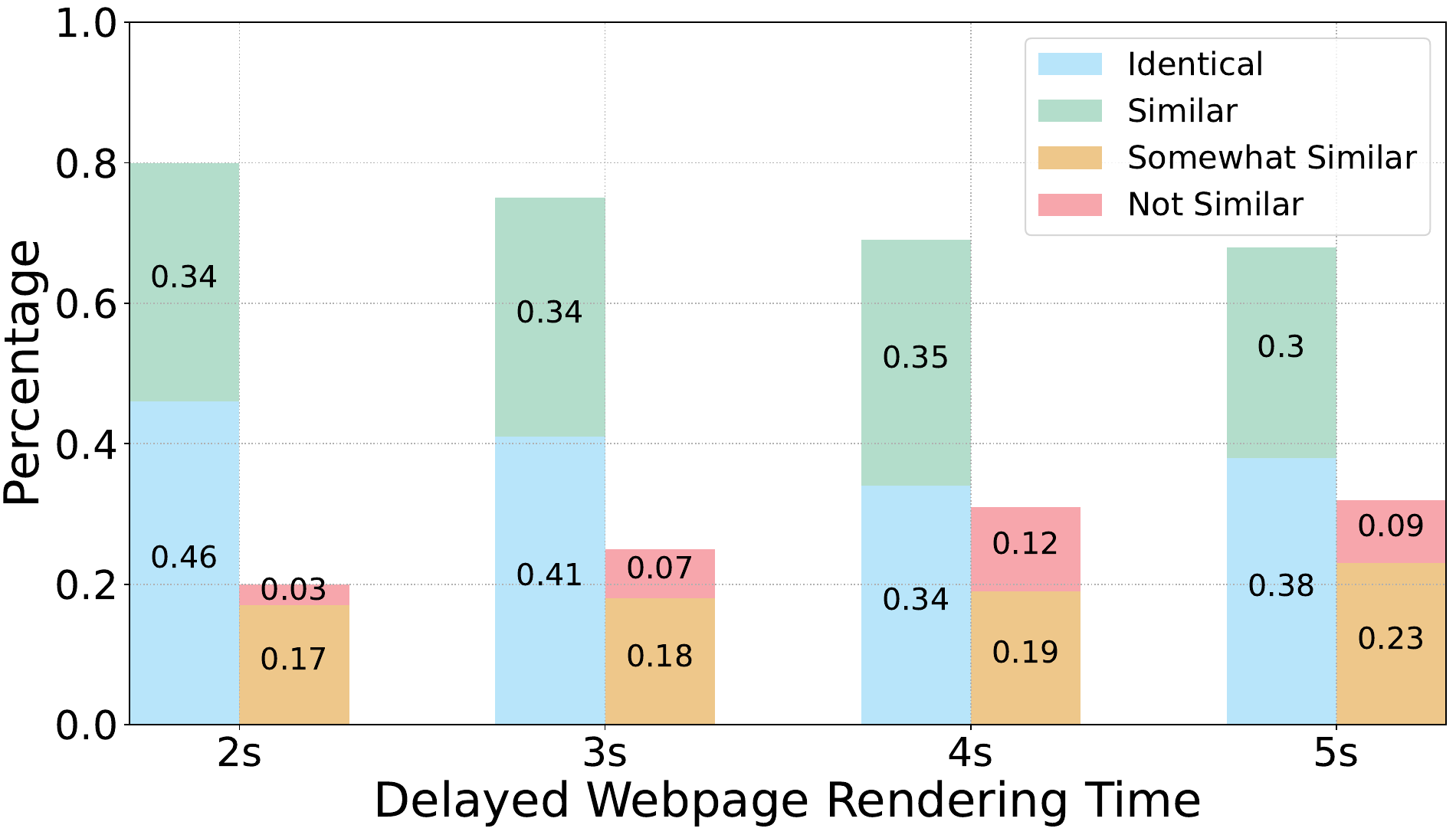}
\vspace{-6mm}
\caption{
\textbf{User Perception of Rendering Delays.} For RQ3a, each video pair shows a webpage rendering at normal speed vs. delayed speed (2 to 5s delay). We report the percentage of participants responses in `similar', `identical', `somewhat similar' and `not similar' for each delay.}
\vspace{-3mm}
\label{fig:similar_result}
\end{figure} 

Overall, 73\% of webpages with delayed rendering were rated as identical or similar to those loading at normal speed, showing that users generally had difficulty noticing our attacks. When looking at responses across different delay times (Fig.~\ref{fig:similar_result}), we found that \textit{users seem to notice the attacks when the delay reached 4 seconds}. At this delay, participants rated more webpage pairs as `similar' rather than `identical', and the proportion of `not similar' responses increased significantly from 0.07 to 0.12 (\smamath{p<.05} and \smamath{\chi^2=5.32}), compared to the responses number at 3s. We also ran a chi-square test to compare the increase in `not similar' responses from 2s to 3s, and found that the rise from 0.03 to 0.07 is also statistically significant (\smamath{p<.05} and \smamath{\chi^2=5.47}). However, the increase in `not similar' responses from 3s and 4s is slightly higher than the increase observed from 2s to 3s. These findings suggest that user suspicion became more noticeable at the 4-second delay, likely due to our attack.

\begin{figure}[t]
\centering
\includegraphics[width=\linewidth]{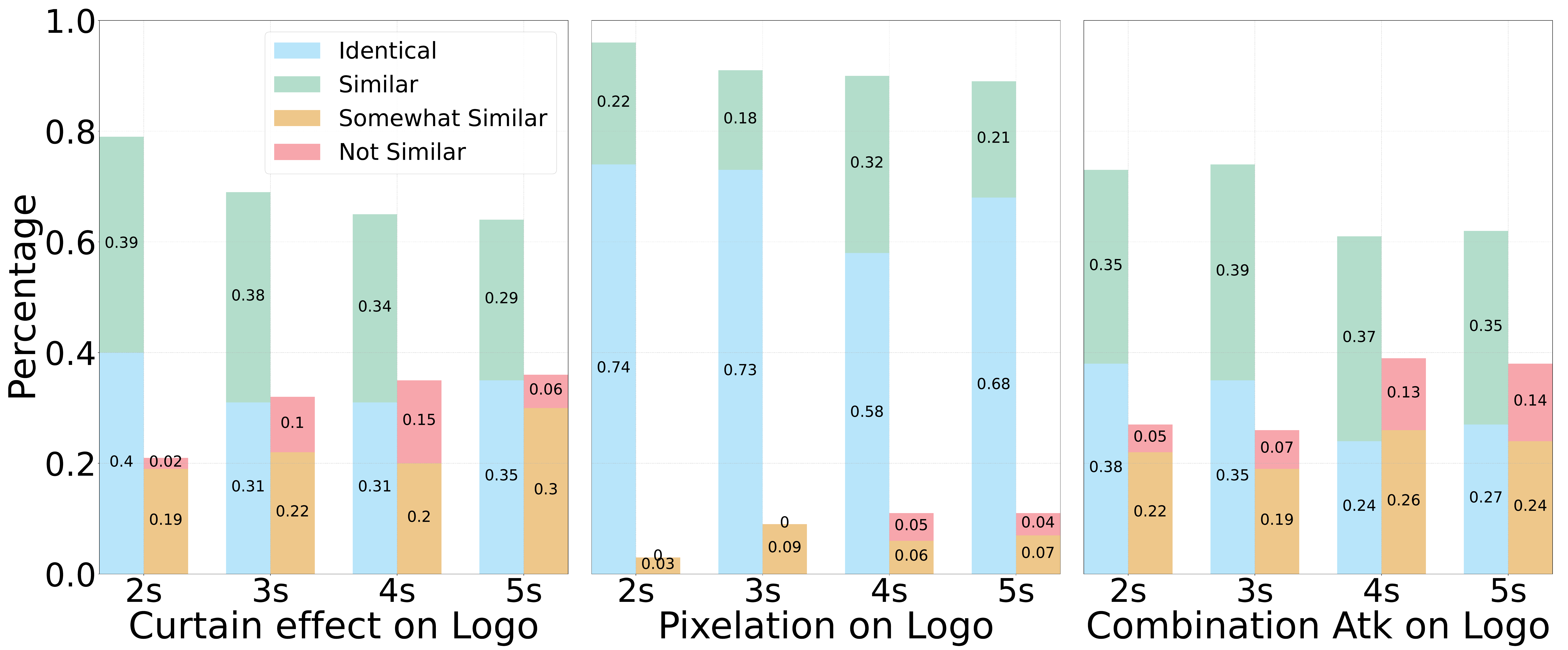}
\vspace{-0.2in}
\caption{
\textbf{User Perception of Different Rendering Attacks.} For RQ3a, each video pair shows a webpage rendering at normal speed vs. delayed speed (2--5s delay). We report the \% of responses in `similar', `identical', `somewhat similar' and `not similar' for for each delay and attack type.
}
\vspace{-0.2in}
\label{fig:similarity_per_atk}
\end{figure}

Moreover, to investigate whether users perceive different attack types at different times, we report their responses across different attacks in Fig.~\ref{fig:similarity_per_atk}. Note that we only examined the perception time for logo-based attacks, as these have the highest impact (both against humans and detectors). For \textit{Curtain effect} attack, the responses in ``not similar" increased from 3s to 4s-delays (0.1 to 0.15). Additionally, the percentage increase of `similar' ratings at 4s also became significant under \textit{Pixelation} attacks, increased from 0.18 to 0.32, compared to 3s. A similar increase appeared in the combination attacks: at 4s, the `not similar' rate increased compared with the 2s and 3s delays (from 0 to 0.05). These findings indicate that participants perceived our delayed rendering attacks when the webpage rendering was delayed to 4 seconds.

\begin{cooltextbox}
    \textbf{\textsc{Answer to RQ3}a:} Participants seem to perceive our attacks when webpages render in 4s (as opposed to 1s).
\end{cooltextbox}

%% file: sections/7-Countermeasures.tex
\section{Countermeasures [RQ4]}
\label{sec:countermeasure}
\noindent 
To answer RQ4, we discuss approaches to mitigate our envisioned attacks (§\ref{ssec:defenses}), and then proposed our own countermeasure (§\ref{ssec:extension}) which we then test (§\ref{ssec:evaluation}).

\subsection{Generic Defenses}
\label{ssec:defenses}
\noindent
Countering delayed-rendering attacks is not trivial.

The most intuitive way to (attempt to) do so is by delaying the capture of the screenshot: for instance, if an attacker forcefully makes logo appear after 5s, a defender may wait 6s. The problem of similar approaches is that it may impair the overall utility of similar detectors for crawlers: phishing websites are generated at a very fast pace~\cite{cic2024phishing}, and waiting for longer timespans (i.e., 3\smamath{\times} or 4\smamath{\times} the default ones) prevents quickly adding newly-found phishing domains to blocklists. Moreover, attackers aware that the capture occurs at a given point in time can simply adjust the rendering behavior so that it falls right outside the capturing window. Even though users may notice it, our user study shows that some users would still be fooled.

Another defense entails signature-based methods that inspect the HTML of a webpage and, if they notice the presence of JavaScript methods that deliberately tamper with the rendering process, flag the webpage as malicious. For instance, developing a signature for our developed \textit{PhishMe} would be trivial; however, foreseeing all possible ways in which delayed-rendering attacks can be implemented is unfeasible.

Finally, phishing education can also help: by raising awareness that attackers can exploit such an apparently natural occurrence  for malicious purposes, users may be more alert when they visit webpages that appear to load less smoothly. However, such efforts not always have the desired degree of effectiveness.~\cite{lain2024content,lain2022phishing}.

\subsection{Proposed Defensive System}
\label{ssec:extension}
\noindent
As a potential countermeasure, we propose a client-side defensive system that protects end-users against phishing websites trying to exploit delayed rendering attacks. 

\subsubsection{Design principles}
\label{sssec:ext_principles}
Given that ``generic'' countermeasures present tradeoffs, we envisage that delayed-rendering attacks can be mitigated by \textit{warning the user that they may have landed on a phishing webpage during their browsing activities.} 

By acknowledging that it is unfeasible to develop a solution that works at scale, it is possible to develop ad-hoc defensive mechanisms that passively analyse the webpages visited by the user, looking for traces of phishing tactics. This can be achieved via browser extensions.

At a high level, such a browser extension must: {\small \textit{(i)}}~operate locally---meaning that it shall not issue queries to external services (to protect user's privacy); {\small \textit{(ii)}}~be customizable and non-invasive~\cite{petelka2019put}, meaning that users should be able to tailor it to their needs, and should avoid drastically blocking a webpage (even when there is reason to believe it is malicious); and {\small \textit{(iii)}}~be lightweight and open source, to foster its adoption and cater to a large audience.

\begin{figure}[!t]
    
    \centering
    \includegraphics[width=0.6\linewidth]{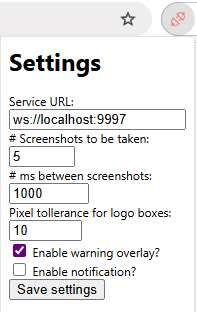}
    \caption{\textbf{User Interface of our browser extension.} Our prototype can be configured to tailor for the users' needs.}
    \vspace{-2mm}
    \label{fig:ExtPopupUI} 
    
\end{figure}

\begin{figure}[!t]
    \centering
    \frame{\includegraphics[width=0.9\linewidth]{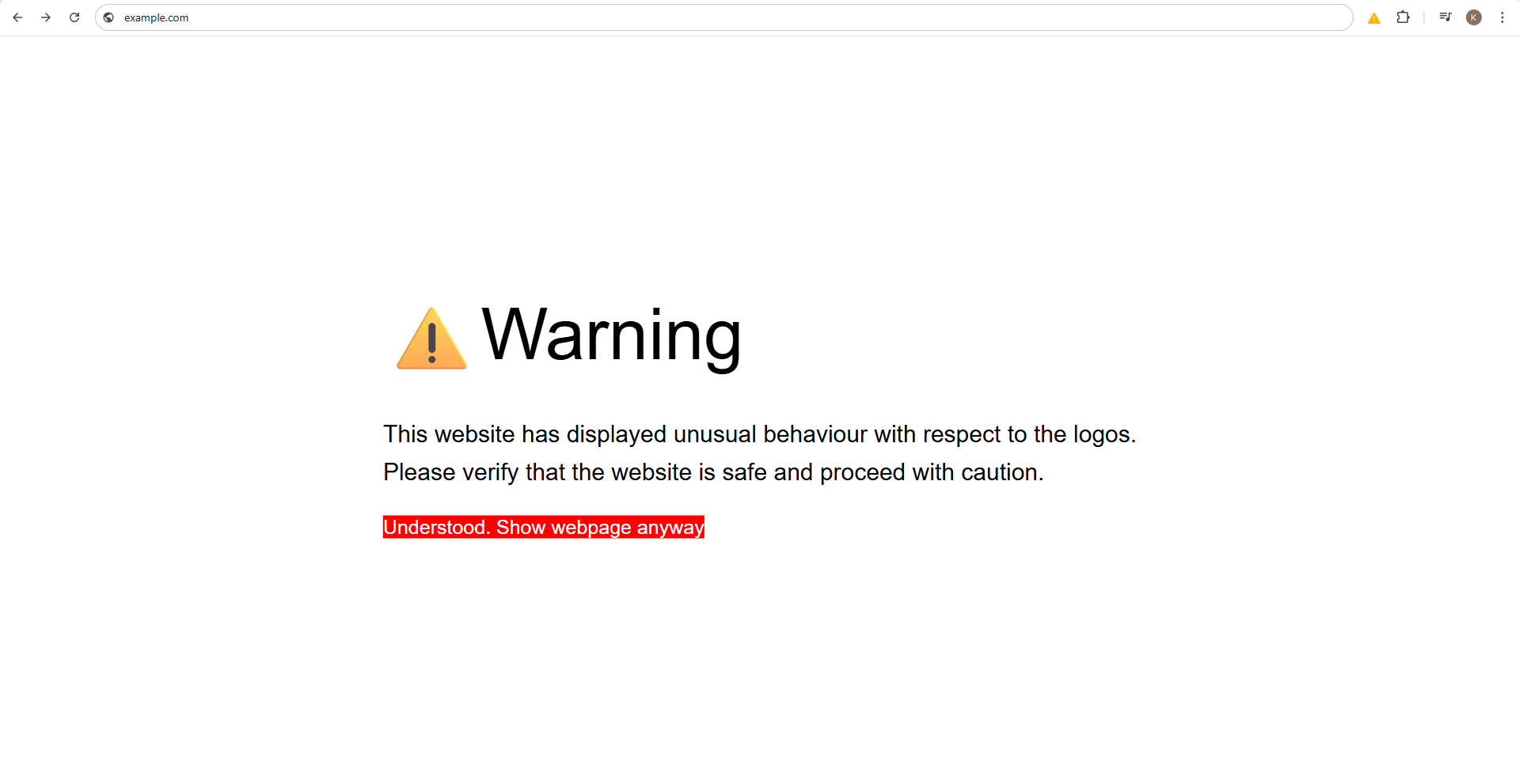}}
    \vspace{-2mm}
    \caption{\textbf{Warning overlay.} If our browser extension detects something suspicious, it will warn the user. }
    \label{fig:ExtOverlay} 
    % \vspace{-4mm}
\end{figure}

\subsubsection{Architecture and Development}
\label{sssec:ext_development}
Our proposed solution is comprised of three main elements: a browser extension that collects data on the visible parts of the active browser tab; a websocket server (which can be deployed anywhere) that handles the communication to and from the extension; a logo detector that identifies the logos contained in a provided screenshot and returns their coordinates. The reason why the logo detector is separate from the extension is due to computational efficiency.

Intuitively, the extension captures screenshots perpetually, and sends them to the logo detector: if such a component determines that the screenshot taken at time \smamath{t+1} contains a logo that was not identified at time \smamath{t}, then this could be indicative of a delayed-rendering attack embedded in the webpage that the user is visiting. In this case, the user should be warned (e.g., with a popup).

We implemented a prototype of the aforementioned solution. For the websocket, we used Python3; for the extension, we considered the Chrome browser (due to its popularity~\cite{hsu2024chrome}): for the logo detector, we used the one in PhishIntention (but without the OCR module). 
We show in Fig.~\ref{fig:ExtPopupUI} the GUI of our extension's prototype, displaying the URL of the websocket server, the total number of screenshot to take whenever a page is visited, and the interval between each capture, as well as a field which regulates the tolerance level for extracting the logos. If a user lands on a suspicious webpage, the extension will trigger the ``warning overlay'' shown in Fig.~\ref{fig:ExtOverlay}. 

\subsubsection{Disclaimers}
\label{sssec:ext_considerations}
We do not claim that our countermeasure is omnipotent---especially in its current implementation. For instance, a known issue of PhishIntention is that it may not recognize the correct logo in a webpage having multiple logos. At the same time, the parameters can provide flexibility but attacker can wait long enough so that the extension does not capture any more screenshots. Moreover, the extension may trigger false positives (e.g., if benign webpages have visual elements which change dynamically), which is why we implemented an ``allow list'' functionality that can be used to prevent the extension from working when landing on webpages known to be benign.
Finally, the logo detector does require some computational power to be run, given that it integrates deep learning models which are supposed to quickly alert if a match is found within the reference list: such a peculiarity may prevent deployment of the websocket server on low-power devices. Nonetheless, our prototype can be expanded to, e.g., include additional ways to detect logos, and the websocket server can be deployed on a powerful machine deployed in a LAN entirely controlled by the device owners, ensuring that all clients (including low-power ones) can benefit from such a solution. 

To the best of our knowledge, we are the first to propose, and practically implement, a ``defensive'' browser extension that protects against delayed-rendering attacks (e.g., the extension in~\cite{varshney2025login} has a different purpose).

\subsection{Performance Evaluation}
\label{ssec:evaluation}
\noindent
We tested our prototype with a proof-of-concept evaluation. First, we checked how much computational resources it requires. We found that deploying the extension and the websocket server on a commodity laptop induces Chrome to increase its RAM utilization by \smamath{\approx}50MB. Of course, the logo detector uses much more resources, but this is not a limitation since it can be deployed anywhere. Then, we checked if our solution could detect our adversarial webpages. To this end, we set the extension to capture five screenshots (one every second). Then, we deployed 24 custom-built adversarial phishing webpages we created for our assessment on a local webserver; we considered one adversarial variant per original phishing webpage. We then visited these webpages with our browser. Out of 24 phishing webpages, 23 correctly trigger a warning. For the 24th, no warning is triggered because of a failure of a component of PhishIntention, which led to a crash; this issue, however, is unrelated to the detection logic of our approach, and (we believe) is attributable to a ``bug'' of PhishIntention's logo-matching process. 

The source code of our entire system is in our repository~\cite{repository}, which also provides a demonstrative video showing the extension flagging phishing webpages that tamper with the rendering process.

%% file: sections/8-discussion.tex
\section{Discussion and Future Work}
\label{sec:discussion}
\noindent
We scrutinise the delayed-rendering attack under a real-world lens~(§\ref{ssec:real}), then we discuss we discuss our findings~(§\ref{ssec:lessons}), and outline limitations (§\ref{ssec:limitations}). % and conclude our work (§\ref{ssec:conclusions})

\subsection{Real-World Considerations}
\label{ssec:real}
\noindent
Do phishers use rendering-based strategies, and do they affect operational detectors? In what follows, we provide evidence supporting the thesis that the answer to both of these questions is ``yes''.

\textbf{Prior work.}
Real attackers are constantly looking for new ways to bypass phishing detectors. In this context, we underscore that \textit{phishers have been exploiting attacks reliant on manipulation of the webpage's rendering for several years}. Indeed, in their S\&P'21 work, Zhang et al.~\cite{zhang2021crawlphish} discuss a large-scale analysis (done in 2019) of ``cloaking'' tactics adopted by phishing webpages, finding that phishing webpages do indeed leverage ``timing'' techniques to, e.g., ``deliberately make rendering slow'', which can lead to evasion. However, visual phishing detectors -- such as Phishintention~\cite{liu2022inferring}, VisualPhishNet~\cite{abdelnabi2020visualphishnet}, and Phishpedia~\cite{lin2021phishpedia} -- became popular \textit{after} the publication of~\cite{zhang2021crawlphish}. We find it surprising that, according to our literature review (in §\ref{ssec:slr}), this vulnerability had not been systematically evaluated insofar, as only limited attention was given by related work to how robust visual phishing detectors can be to rendering-based evasion strategies.

\textbf{In-the-wild evidence.}
Setting aside the 2019 findings by Zhang et al.~\cite{zhang2021crawlphish}, we provide additional evidence that rendering-based evasion strategies very likely \textit{affect operational visual phishing detectors}. The authors of a SaTML'23 paper~\cite{apruzzese2023real} examined the failures of an industrial phishing website detector reliant on deep learning. Alongside the case study discussed in the paper, however, the authors also released a dataset containing the screenshots of 100 phishing webpages which evaded the detection. By inspecting this dataset, we found that several of these screenshots were captured before the page could fully render (e.g., sample \#58 of this dataset shows a webpage mimicking a Microsoft's login form, but the background image rendered only partially; the same can be said for, e.g., \#43 which mimicks Chase bank). Even though we cannot claim with certainty that the reason for the misclassification was solely due to rendering issues, it is factual that similar strategies are still employed by phishers. This finding inspired us to carry out the systematic assessment described in this paper.

\textbf{Google Chrome.} Our empirical evaluation (in §\ref{sec:results}) predominantly focused on research prototypes. However, we have also tested, in a black-box fashion, the client-side detector of Google Safe Browsing~\cite{gsb} integrated in Google Chrome (which allegedly also relies on visual similarity~\cite{miao2023good}). To this end, we host all our considered phishing webpages on a (local) HTTP server. We customise our webpages so that they load a different variant of our delayed-rendering attack every 1.5 seconds. We then use Google Chrome (v146) to visit these phishing pages, manually checking if there is any warning/alert raised by the browser. Despite remaining $>$30 seconds on each webpage ($\approx$90s for pages with background, and $\approx$31s for those without), no suspicious behavior is detected by the browser. We also visited them with Chromium (v149), and even in this case, no warning was triggered.

\subsection{Major Findings and Lessons Learned}
\label{ssec:lessons}
\noindent
We examined an overlooked class of phishing attacks that is designed to bypass visual phishing detectors.

First, our evaluation showed that state-of-the-art systems can be trivially evaded by combining delayed-rendering techniques such as pixelation and curtain effects (§\ref{ssec:main_test}). Given that human users are also deceived (§\ref{ssec:survey_result}), lack of research on this subtle issue (which we highlighted in §\ref{ssec:slr}) is worrying. We endorse future work to further investigate the effectiveness of delayed-rendering attacks. For instance, other visual detectors can be assessed, other forms of perturbations can be explored, potentially applied to different webpages. 

Second, it is factual that countermeasures that work ``at scale'' are hard to devise. However, we underscore that delayed-rendering attacks can target also other classes of phishing detectors: for instance, SpacePhish~\cite{spacephish2022} performs its detection by computing features based on the HTML of a webpage---which can dynamically change, and so there is the risk that the moment when the features are computed does not reflect the actual intention of the webpage. We therefore advocate for the development of defenses specifically designed to withstand similar attacks.

Third, delayed-rendering attacks are complementary to other forms of ``adversarial ML attacks'' proposed in prior work. Our threat model does not differ substantially from, e.g., the one envisioned in~\cite{hao2024doesn}. However, we emphasize that the attack proposed in~\cite{hao2024doesn} is much more expensive, since it requires to explore a large space of possible logos that can fool a surrogate classifier. In contrast, the attack proposed in this work simply require to delay the rendering of certain elements so that they load in \smamath{\approx}4s (to avoid alerting users---who would still be deceived even if the delay is higher). Regardless, our attack can also be combined with that proposed in~\cite{hao2024doesn}.

\subsection{Limitations}
\label{ssec:limitations}
\noindent
We do not see threats to the validity of our conclusions.

Our paper is mostly ``U.S. centric''. We considered brands popular in the U.S. (as also done in most prior work~\cite{lin2021phishpedia, liu2022inferring, hao2024doesn}). We deliberately aligned our user study to such a setup by ensuring that only U.S. people participated. Yet, we acknowledge that phishing is endemic also in other areas~\cite{gopavaram2021cross,yuan2025beyond} of the world---albeit delayed-rendering attacks have no notion of geographical location.

Our evaluation encompasses pixelation and curtain effects, applied to dozens of webpages. Yet, our results are consistent, i.e., expanding our evaluation by considering a larger dataset would not change our conclusions. Furthermore, as we remarked (§\ref{ssec:gap}) exploring all possible ways to convey delayed-rendering attacks is unfeasible.

We acknowledge that the potential for delayed-rendering attacks was mentioned in some prior work (§\ref{ssec:slr}). However, we are the first to practically explore the impact of such a threat, which we also validate with an original user study (considering both a primed and non-primed setup), and which we attempt to mitigate by proposing and developing an open-source solution reliant on browser extentions. We posit that showing the simplicity, and effectiveness, of such attacks is a substantial scientific contribution to the security community.

%% file: sections/9-conclusions.tex
\section{Conclusions}
\label{sec:conclusions}
\noindent
We cast light on a security vulnerability that affects anti-phishing schemes used also by real-world companies. We show that state-of-the-art detectors, including those based on LLMs, can be easily bypassed by delaying the rendering of key visual elements (e.g., the brand logo). An original user study confirms that humans are also susceptible to such an adversarial tactic.

%% file: sections/11-acknowledgment.tex
\section*{Acknowledgments}
\label{ack}

We would like to thank: the anonymous reviewers of EuroS\&P’26 for their constructive comments and suggestions; and the Anti-Phishing Working Group (APWG) for providing access to phishing URLs used in this study. This work was partially supported by the AI-Sec Lab initiative between Sapienza University of Rome and the CYBER 4.0 Competence Center. This work was partially supported by the Wallenberg AI, Autonomous Systems
and Software Program (WASP). Parts of this research are funded by Hilti. The opinions and findings in this paper do not reflect the views of the sponsors.

%% file: main.bbl

%% file: appendix/structure_app.tex
\input{appendix/openscience}
\input{appendix/harm}
\input{appendix/A-tables}
\input{appendix/B-extend}

%% file: appendix/openscience.tex
\section*{Open Science Expectations}
\label{app:open}
\noindent
We have already clearly stated that our resources will be openly shared (see, e.g., §\ref{sec:introduction}).

For completeness, we have created a repository containing all relevant artifacts produced in the course of this research. The link is: \url{https://github.com/joanyy/eurosp26_delayphish}. Within the repository, we provide: our phishing website dataset, source code (including that of the browser extension), and user study questionnaire.

Please note that our phishing dataset was collected using URLs from APWG, and we do not have permission to disclose the original URLs, but they are publicly available for members. And, we provide the HTML files, screenshots, and components that we collected ourselves, which are the exact materials used in our experiments and are sufficient for reproducing our results.  Additionally, according to our data management policy, we cannot disclose the original user responses collected from Prolific to protect participants privacy. However, we provide our complete user study design to ensure reproducibility.

%% file: appendix/harm.tex
\section*{Proactive Prevention of Harm}
\label{app:harm}
\noindent \textbf{Potential harm during study.} In the user study, participants could potentially be deceived by phishing websites and enter personal credentials, leading to financial losses. To mitigate this risk, we displayed websites through videos rather than live online websites, so that participants did not directly interact with actual phishing websites and were not exposed to any harm.

\textbf{Potential harm from publication.} The proposed attacks could be misused by attackers to bypass current phishing detection systems, which could increase threats and cause financial losses. To mitigate this risk, we propose a defense system in §\ref{sec:countermeasure} that can effectively counter this threat. However, as remarked in §\ref{ssec:real}, attackers are (very likely) already exploiting this vulnerability, so we do not believe that publication of this paper would lead to harm. On the contrary, out paper's findings should serve as a call to action to find ad-hoc solutions to this problem.

%% file: appendix/A-tables.tex
\section{Tables}
\label{app:first}
\noindent
We show the distribution of our dataset in Table~\ref{tab:dataset}, which includes 24 phishing websites. The effectiveness of attacks (curtain effect, pixelation, curtain effect+pixelation) on visual phishing detectors is reported in Table~\ref{tab:pixelation_asr},\ref{tab:combine_asr}, \ref{tab:curtain_asr}, measured by FNR. We provide the demographics of our our study partcipants in Table~\ref{tab:demographic_accuracy}, and the distribution of videos shown in the step (II) and (III) of our user study in Table~\ref{tab:video_p1} and \ref{tab:video_p2}, respectively.

\begin{table}[t]
\caption{\textbf{Webpage distribution in our dataset.} For each brand, we report the number of phishing webpages, whether the samples have background images, and whether the brand is used in user study. (note: for Instagram, one webpage does not have background image)} \label{tab:dataset} 
% \vspace{-2mm}
\small
    \centering
    \resizebox{\linewidth}{!}{%
        \begin{tabular}{c|c|c|c}
            \toprule
            \textbf{Brand} & \textbf{Number} & \textbf{Background} & \textbf{User study}\\
            \midrule
            Outlook&1&N&Y\\
            DHL&1&Y&Y\\
            Wells Fargo&2&Y&Y\\
            Capital One&1&N&N\\
            American Express&1&N&N\\
            Comcast&1&Y&Y\\
            BT&1&Y&N\\
            Alibaba&1&Y&N\\
            Swisscom&1&N&N\\
            PayPal&2&N&Y\\
            Facebook&3&N&Y\\
            Netflix&1&Y&Y\\
            IONOS&1&N&N\\
            Spotify&1&N&Y\\
            Instagram&3&Y&Y\\
            Microsoft&1&Y&N\\
            Yahoo&1&N&N\\
            eBay&1&N&Y\\
            \bottomrule
        
        \end{tabular} 
        }     
\end{table}

\begin{table}[t]
\caption{\textbf{Detectors Baseline Performance.} We report the True Positive Rate and False Negatives of visual phishing detectors on 24 original phishing websites, as their baseline performance.} 
\label{tab:baseline} 
% \vspace{-3mm}
\small
    \centering
    
        \begin{tabular}{c|c|c}
            \toprule
            \textbf{Detector} & $FN$& \textbf{$TPR$}\\
            \midrule
            PhishIntention&0&1\\
            PhishPedia&7&0.708\\
            PhishLLM&2&0.917\\
            VisualPhishNet&9&0.625\\
        \bottomrule
        \end{tabular}     
\end{table}        

\begin{table}[t]
\caption{\textbf{\textit{Curtain effect} effectiveness.} We report the FNR of four visual phishing detectors--PhishLLM, PhishPedia, PhishIntention and VisualPhishNet--under \textit{Curtain effect} attacks, where perturbations are applied to logos, backgrounds or both.
Higher rates indicate more effective attacks (the most effective shown in bold). For example, all adversarial samples bypassed PhishIntention when logo images were rendered with 0\% or 25\% visibility. The `no-atk' row shows the baseline performance of detectors on the 24 original phishing pages.} 
\label{tab:curtain_asr} 
% \vspace{-3mm}
\small
    \centering
    \resizebox{\linewidth}{!}{%
        \begin{tabular}{c|c|c|c|c|c}
            \toprule
            \textbf{Atk. intensity}&\textbf{Attack target}&\textbf{PhishIntention} & \textbf{PhishPedia} & \textbf{PhishLLM} & \textbf{VisualPhishNet}\\
            \hline
            \multirow{3}{*}{Invisible} & logo&\textbf{1.0}&\textbf{1.0}&0.958&0.292\\ \cline{2-6}
           & background&0.0&0.3&0.1&0.4 \\ \cline{2-6}
           
           & logo \& background & \textbf{1.0}&\textbf{1.0}&\textbf{1.0}&\textbf{0.6}\\ \hline
            \multirow{3}{*}{25\%} & logo&\textbf{1.0}&0.958&0.792&0.25\\ \cline{2-6}
            
           & background&0.0&0.3&0.2&0.5 \\ \cline{2-6}
           & logo \& background &\textbf{1.0}&0.9&0.8&0.4  \\ \hline
            \multirow{3}{*}{50\%} & logo&0.917&0.875&0.333&0.25\\ \cline{2-6}
            
           & background&0.0&0.3&0.2&0.2 \\ \cline{2-6}
           & logo \& background & 0.9&0.7&0.3&0.2 \\ \hline         \multirow{3}{*}{75\%} & logo&0.292&0.5&0.25&0.292\\ \cline{2-6}
           
           & background&0.0&0.3&0.0&0.1\\ \cline{2-6}
           & logo \& background & 0.1&0.3&0.1&0.1 \\ \cline{2-6} 
           \hline
           no-atk&baseline& 0&0.292&0.083&0.375\\
            \bottomrule
        \end{tabular} 
        }
         
% \vspace{-0.1in}
\end{table}

\begin{table}[t]
\caption{\textbf{\textit{Pixelation} effectiveness.} FNR of four visual phishing detectors under \textit{Pixelation} attacks. } 
\label{tab:pixelation_asr} 
% \vspace{-2mm}
\small
    \centering
    \resizebox{\linewidth}{!}{%
        \begin{tabular}{c|c|c|c|c|c}
            \toprule
            \textbf{Atk. intensity}&\textbf{Attack target}&\textbf{PhishIntention} & \textbf{PhishPedia} & \textbf{PhishLLM} & \textbf{VisualPhishNet}\\
            \hline
            \multirow{3}{*}{5x5px} & logo&0.625&0.708&0.167&0.375\\ \cline{2-6}
           & background& 0.0&0.3&0.0&0.1\\ \cline{2-6}
           
           & logo \& background &\textbf{0.7}&\textbf{0.8}&0.1&0.1  \\ \hline
            \multirow{3}{*}{4x4px} & logo&0.5&0.588&\textbf{0.208}&0.375\\ \cline{2-6}
            
           & background&0.0&0.3&0.1&0.1 \\ \cline{2-6}
           & logo \& background &\textbf{0.7}&0.7&0.0&0.1  \\ \hline
            \multirow{3}{*}{3x3px} & logo&0.458&0.5&0.167&\textbf{0.417}\\ \cline{2-6}
            
           & background&0.0&0.3&0.2&0.1\\ \cline{2-6}
           & logo \& background &0.5&0.5&0.1&0.3 \\ \hline         \multirow{3}{*}{2x2px} & logo&0.25&0.375&0.167&\textbf{0.417}\\ \cline{2-6}
           
           & background&0.0&0.3&0.1&0.1\\ \cline{2-6}
           & logo \& background &0.2&0.4&0.0&0.3 \\ \hline
           no-atk&baseline& 0&0.292&0.083&0.375\\
           % \cline{2-6} 
            \bottomrule
        \end{tabular} 
        }

\end{table}

\begin{table}[t]
\caption{We report the distribution of videos used in the user study for RQ3a. The videos depict webpage rendering under different delay times and attacks.}
\label{tab:video_p1} 
% \vspace{-3mm}
\small
    \centering
    \resizebox{\linewidth}{!}{%
        \begin{tabular}{c|c|c|c|c|c}
            \toprule
            \textbf{Attack target} & \textbf{Attacks} & \textbf{2s} & \textbf{3s} & \textbf{4s}&\textbf{5s}\\
            \midrule
            Logo & Curtain effect&126&124&110&134\\           
           Logo & Pixelation& 58&66&66&57\\ 
           Logo & Combination &118&127&124&125 \\
       \bottomrule
        \end{tabular} 
        }  

\end{table}

\begin{table}[t]
\caption{\textbf{Combination attacks effectiveness.} The FNR of four visual phishing detectors under combined attacks (i.e., apply \textit{Pixelation} and \textit{Curtain effect} attacks simultaneously).} 
\label{tab:combine_asr} 
% \vspace{-3mm}
\small
    \centering
    \resizebox{\linewidth}{!}{%
        \begin{tabular}{c|c|c|c|c|c}
            \toprule
            \textbf{Atk. intensity}&\textbf{Attack target}&\textbf{PhishIntention} & \textbf{PhishPedia} & \textbf{PhishLLM} & \textbf{VisualPhishNet}\\
            \hline
            \multirow{3}{*}{5x5px \& 25\%} &
            logo&\textbf{1.0}&\textbf{1.0}&0.917&0.25\\ \cline{2-6}
           & background&0.0&0.3&0.2&0.3 \\ \cline{2-6}
           & logo \& background &\textbf{1.0}&\textbf{1.0}&\textbf{1.0}&\textbf{0.5} \\ \hline
           \multirow{3}{*}{5x5px \& 50\%} &
            logo&0.917&0.958&0.958&0.292\\ \cline{2-6}
           & background& 0.0&0.3&0.2&0.2\\ \cline{2-6}
           & logo \& background & 0.9&0.9&0.5&0.3\\ \hline
            \multirow{3}{*}{5x5px \& 75\%} &
            logo&0.708&0.833&0.125&0.292\\ \cline{2-6}
           & background& 0.0&0.3&0.2&0.1\\ \cline{2-6}
           & logo \& background & 0.7&0.8&0.2&0.1\\ \hline
            \multirow{3}{*}{4x4px \& 25\%} &
            logo&\textbf{1.0}&\textbf{1.0}&0.875&0.25\\ \cline{2-6}
           & background&0.0&0.3&0.0&0.4 \\ \cline{2-6}
           & logo \& background &\textbf{1.0}&\textbf{1.0}&0.8&\textbf{0.5} \\ \hline
           \multirow{3}{*}{4x4px \& 50\%} &
            logo&0.917&0.917&0.458&0.25\\ \cline{2-6}
           & background&0.0&0.3&0.1&0.2 \\ \cline{2-6}
           & logo \& background &0.9&0.8&0.5&0.3 \\ \hline
            \multirow{3}{*}{4x4px \& 75\%} &
            logo&0.625&0.75&0.125&0.333\\ \cline{2-6}
           & background&0.0&0.3&0.2&0.1 \\ \cline{2-6}
           & logo \& background & 0.7&0.7&0.0&0.1\\ \hline

          \multirow{3}{*}{3x3px \& 25\%} &
            logo&\textbf{1.0}&\textbf{1.0}&0.875&0.25\\ \cline{2-6}
           & background& 0.0&0.3&0.2&0.4\\ \cline{2-6}
           & logo \& background &\textbf{1.0}&\textbf{1.0}&\textbf{1.0}&\textbf{0.5} \\ \hline
           \multirow{3}{*}{3x3px \& 50\%} &
            logo&0.917&0.875&0.333&0.25\\ \cline{2-6}
           & background&0.0&0.3&0&0.2 \\ \cline{2-6}
           & logo \& background &0.9&0.7&0.4&0.3 \\ \hline
            \multirow{3}{*}{3x3px \& 75\%} &
            logo&0.625&0.667&0.125&0.333\\ \cline{2-6}
           & background&0.0&0.3&0.1&0.1 \\ \cline{2-6}
           & logo \& background & 0.7&0.5&0.2&0.1\\ \hline
     
            \multirow{3}{*}{2x2px \& 25\%} &
            logo&\textbf{1.0}&0.958&0.833&0.25\\ \cline{2-6}
           & background&0.0&0.3&0.2&0.4 \\ \cline{2-6}
           & logo \& background & \textbf{1.0}&0.9&0.8&\textbf{0.5}\\ \hline
           \multirow{3}{*}{2x2px \& 50\%} &
            logo&0.917&0.875&0.375&0.25\\ \cline{2-6}
           & background& 0.0&0.3&0.1&0.2\\ \cline{2-6}
           & logo \& background &0.9&0.7&0.3&0.2 \\ \hline
            \multirow{3}{*}{2x2px \& 75\%} &
            logo&0.625&0.625&0.25&0.292\\ \cline{2-6}
           & background& 0.0&0.3&0.2&0.1 \\ \cline{2-6}
           & logo \& background &0.5&0.5&0.2&0.1 \\ \hline
        no-atk&baseline& 0&0.292&0.083&0.375\\
 
            \bottomrule
        \end{tabular} 
        }
        
\end{table}

\begin{figure}[t]
 \centering
\subfloat[Original phishing webpage]{\label{fig:visual_a}
\includegraphics[width=0.45\linewidth]{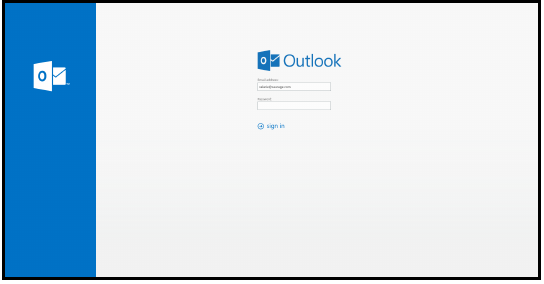}
}
\hfill
\subfloat[Matched Target (distance: 2.1)]{\label{fig:b}
\includegraphics[width=0.45\linewidth]{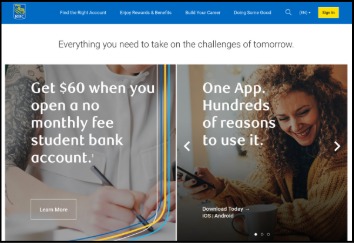}
}
\\
\subfloat[Adv. webpage (Logo Invisible)]{\label{fig:visual_c}
\includegraphics[width=0.45\linewidth]{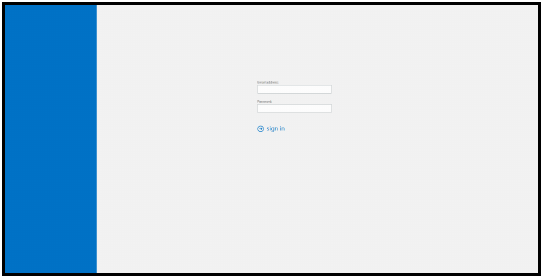}
}
\hfill
\subfloat[Matched Target (distance:0.73)]{\label{fig:visual_d}
\includegraphics[width=0.45\linewidth]{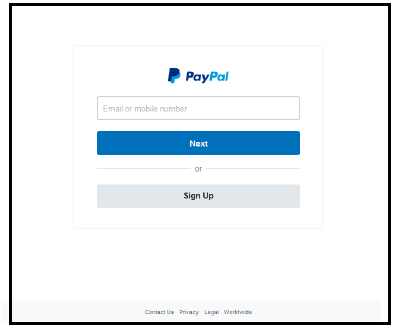}
}
\caption{
\textbf{Assessment on VisualPhishNet.}%
   { (a) Original Phishing webpage screenshot (Outlook) and (b) its target identified by VisualPhishNet. Their visual distance of 2.1 exceeds the threshold, resulting in misclassification as Benign. (c) Webpage after applying \textit{Curtain Effect} attack to conceal the logo. VisualPhishNet identifies PayPal as the target (d) with visual distance 0.73, below the threshold, correctly flagging it as Phish.} 
}
% \vspace{-0.2in}
\label{fig:visual_samples}
\end{figure}

\begin{table}[t] 
\caption{\textbf{Demographics of our user study (in §\ref{sec:userstudy})}.}
    \label{tab:demographic_accuracy}
    % \vspace{-3mm}
\centering 
%[H]
\small 
\resizebox{\columnwidth}{!}{
\begin{tabular}{l c} 
\toprule
\textbf{User Study} & \textbf{\# of Participants}\\  
\midrule  
\multicolumn{2}{l}{\textbf{Gender}}\\
\hspace{3mm}{Male} & {122}\\ 
\hspace{3mm}{Female} & 122\\
\hspace{3mm}{Non-binary / third gender} & {1}\\ 
\hspace{3mm}{Prefer not to say} & {2}\\
\midrule  
\multicolumn{2}{l}{\textbf{Age}}\\
\hspace{3mm}{18-29} & 29 \\ 
\hspace{3mm}{30-39} & 76 \\ 
\hspace{3mm}{40-49} & 60 \\ 
\hspace{3mm}{50-59} & 49\\ 
\hspace{3mm}{60-69} &26\\ 
\hspace{3mm}{70 or above} &5\\ 
\hspace{3mm}{Prefer not to say} & 2\\ 
  \midrule 
\multicolumn{2}{l}{\textbf{Education}}\\
\hspace{3mm}{Some high school or less} &5\\  
\hspace{3mm}{High school diploma or GED} &34\\ 
\hspace{3mm}{Some college, but no degree} &35 \\ 
\hspace{3mm}{Associates or technical degree} &28\\ 
\hspace{3mm}{Bachelor’s degree} & 85\\ 
\hspace{3mm}{Graduate or professional degree (MA, MS, MBA, PhD, JD, MD, DDS etc.)} &58\\
\hspace{3mm}{Prefer not to say} &2\\  

\midrule 
\multicolumn{2}{l}{\textbf{Technical Background in Computer Science/Engineering}}\\
\hspace{3mm}{Yes} & 50\\  
\hspace{3mm}{No} & 192\\
\hspace{3mm}{Prefer not to say} &5\\
\midrule  

\multicolumn{2}{l}{\textbf{Technical Background in Cybersecurity}}\\
\hspace{3mm}{Yes} &  22  \\  
\hspace{3mm}{No} &221  \\
\hspace{3mm}{Prefer not to say} &4  \\ 
  \bottomrule  
\end{tabular}
}

\end{table}

\begin{table}[t]
\caption{We report the distribution of videos used in the user study for RQ3b. The videos depict webpage rendering under different delay times and attacks.}
    
        \label{tab:video_p2} 
        % \vspace{-3mm}
\small
    \centering
    \resizebox{\linewidth}{!}{%
        \begin{tabular}{c|c|c|c|c|c}
            \toprule
            \textbf{Attack target} &\textbf{Attacks}& \textbf{2s} & \textbf{3s} & \textbf{4s}&\textbf{5s}\\
            \midrule

            \multirow{3}{*}{Logo} & Curtain effect&181&192&186&182\\ \cline{2-6}
           & Pixelation& 181&187&169&204\\ \cline{2-6}
           & Combination &193&190&163&195 \\ \hline
           
        \multirow{3}{*}{Background} & Curtain effect&124&121&123&126\\ \cline{2-6}
           & Pixelation&117&123&127&127 \\ \cline{2-6}
           & Combination &174&192&183&192 \\ \hline

        \multirow{3}{*}{Combination} & Curtain effect&186&201&174&180\\ \cline{2-6}
           & Pixelation&177&188&195&181  \\ \cline{2-6}
           & Combination &177&191&180&193 \\
            \bottomrule
        \end{tabular} 
        }

% \vspace{-0.2in}
\end{table}

\vspace{0.1in}

%% file: appendix/B-extend.tex
\section{Extended Experiments}
\label{app:extend}

\noindent 
To further examine the broad applicability of delayed-rendering attack strategies, we expanded our evaluation to encompass more webpages/brands and, particularly, see if our adversarial strategies do not backfire. In other words, we wonder (RQ2a):~``\textit{\purple{Do webpages that already evade Phishintention still evade it after applying our delayed-rendering attack?}}'' 

\textbf{Approach.}
We randomly sampled 18 samples from the 866 phishing websites that were misclassified as benign by PhishIntention, covering 13 different brands\footnote{These samples are different from the 24 webpages that PhishIntention correctly identified as phishing in our main evaluation~(§\ref{sec:results}).}. Among these, 11 websites contain background images, while 7 do not. 
We then applied our curtain effect and pixelation attacks to generate 800 adversarial samples. To answer RQ2a, we submit these adversarial webpages to Phishintention. Finally, and as a complementary assessment, we also send these webpages to the three other detectors considered in our evaluation, i.e., VisualPhishNet~\cite{abdelnabi2020visualphishnet}, PhishPedia~\cite{lin2021phishpedia}, and PhishLLM~\cite{liu2024less}.

\begin{figure}[!htbp]
\centering
\includegraphics[width=\linewidth]{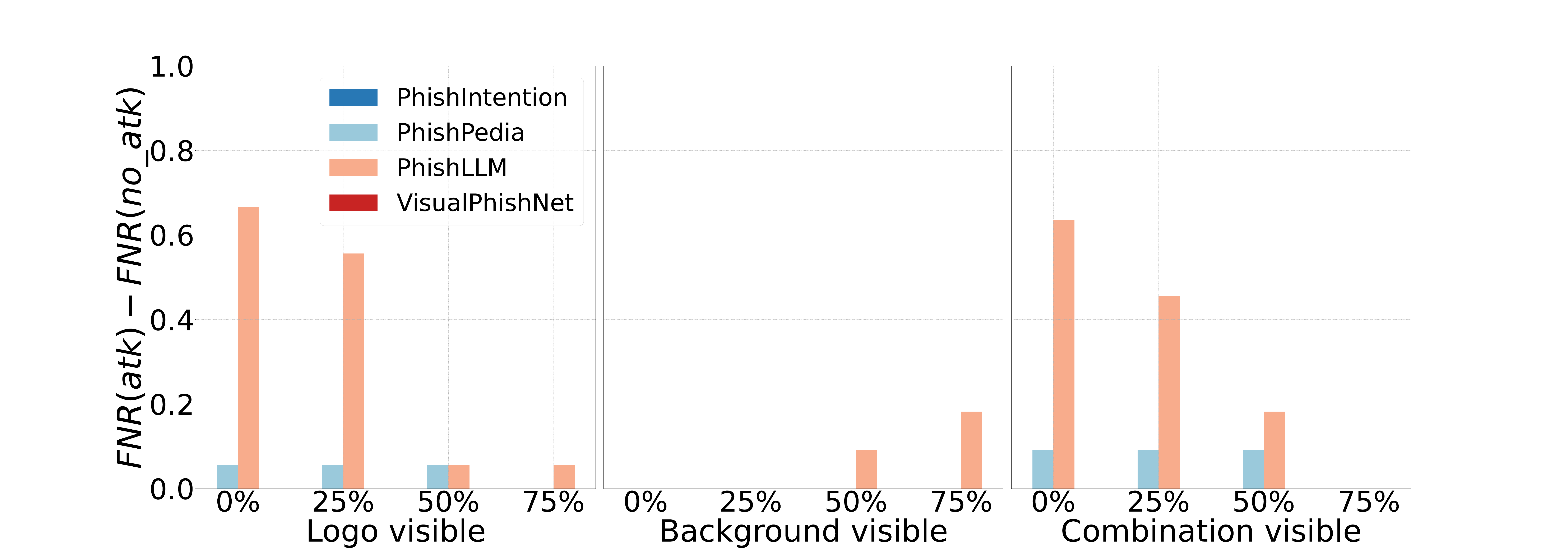}
% \vspace{-5mm}
\caption{
\textbf{Curtain effect effectiveness (extended).} We report the increase in FNR of four visual phishing detectors when Curtain effect is applied to 18 phishing webpages, compared to the no-atk baseline. The phishing website screenshots are captured during the rendering process of logo, background, or both images, with visibility ranging from 0\% to 75\%.}
% \vspace{-4mm}
\label{fig:curtain_extend}
\end{figure}

\begin{figure}[!htbp]
\centering
\includegraphics[width=\linewidth]{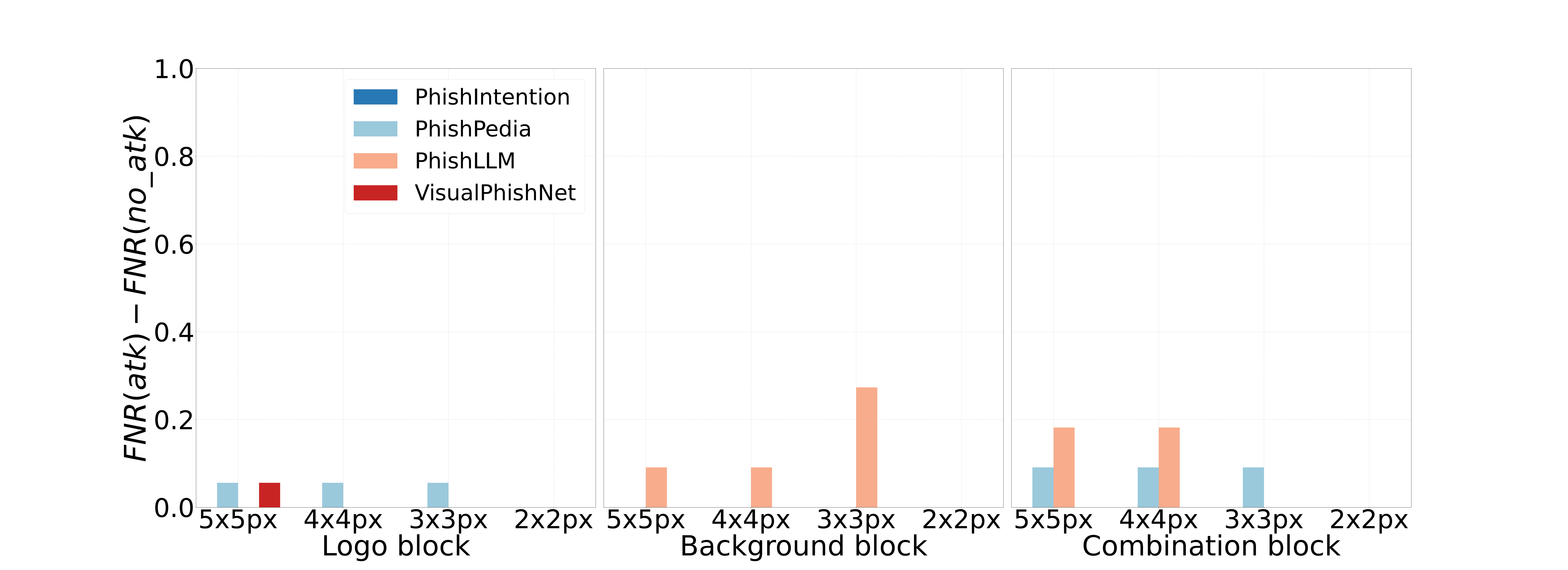}
% \vspace{-0.3in}
\caption{
\textbf{Pixelation effectiveness (extended).} We report the increase in FNR of four visual phishing detectors when \textit{Pixelation} is applied to 18 phishing webpages, compared to the no-atk baseline. These adversarial screenshots are captured during the rendering of logo images, background images, or both, with pixelation block sizes ranging from 2×2 to 5×5 pixels.} 
% \vspace{-0.1in}
\label{fig:pixelation_extend}
\end{figure}

\textbf{Results.}
The results shown in Fig.~\ref{fig:curtain_extend} and Fig.~\ref{fig:pixelation_extend}, reported as an increase in FNR under different attacks, compared to the baseline $FNR (no\_atk)$, indicate that most of our attacks consistently enhance evasion success across different phishing detectors. For example, when applying the curtain effect attack to the logos of these phishing samples, the FNR of PhishLLM consistently increased by 0.67 as logo visibility decreased from 100\% to 0\% (see the subfigure `Logo visibe' of Fig.~\ref{fig:curtain_extend}), reaching an evasion rate of 94\%. For PhishIntention, the FNR increase is 0, since its FNR without any attack is already 100\% (i.e., fully evaded the detector), leaving no room for further improvement: therefore, we can answer RQ2a with a clear ``yes''. However, we observe a few perturbations slightly reduce evasion effectiveness \textit{against other detectors}. For example, phishing pages with the curtain effect attack show a small drop in FNR against VisualPhishNet, from 0.778 to 0.667 (see Table~\ref{tab:curtain_fnr}). This happens because VisualPhishNet compares the visual similarity of the entire page. When the key identifying element, the logo, is hidden, the detector shifts its focus to other parts of the page. The remaining layout may closely match a different known target, causing the detector to misclassify which target is being impersonated. However, it still correctly identifies the page as phishing.

\begin{table}[!htbp]
\caption{\textbf{\textit{Curtain effect} effectiveness (extended).} FNR of four visual phishing detectors under \textit{Curtain effect} attacks. The `no-atk' row shows the baseline performance of these detectors on the 18 original phishing webpages (unperturbed).} 
\label{tab:curtain_fnr} 
% \vspace{-3mm}
\small
    \centering
    \resizebox{\linewidth}{!}{%
        \begin{tabular}{c|c|c|c|c|c}
            \toprule
            \textbf{Atk. intensity}&\textbf{Attack target}&\textbf{PhishIntention} & \textbf{PhishPedia} & \textbf{PhishLLM} & \textbf{VisualPhishNet}\\
            \hline
            \multirow{3}{*}{Invisible} & logo& 1.0&1.0&0.944&0.667\\ \cline{2-6}
           & background& 1.0&0.909 &0.182&0.727  \\ \cline{2-6}
           
           & logo \& background &1.0&1.0&0.909 &0.545   \\ \hline
            \multirow{3}{*}{25\%} & logo&1.0&1.0&0.833 &0.556  \\ \cline{2-6}
            
           & background& 1.0&0.909 &0.273&0.636   \\ \cline{2-6}
           & logo \& background & 1.0&1.0&0.727 &0.364  \\ \hline
            \multirow{3}{*}{50\%} & logo&1.0&1.0&0.33&0.722  \\ \cline{2-6}
            
           & background&1.0&0.909 &0.364&0.636   \\ \cline{2-6}
           & logo \& background & 1.0&1.0&0.455&0.636   \\ \hline         \multirow{3}{*}{75\%} & logo& 1.0&0.944&0.33 &0.667 \\ \cline{2-6}
           
           & background& 1.0&0.909 &0.455&0.727  \\ \cline{2-6}
           & logo \& background & 1.0&0.909&0.273&0.636   \\ \cline{2-6} 
           \hline
           no-atk&baseline&1.0&0.944&0.278&0.778 \\
            \bottomrule
        \end{tabular} 
        }

% \vspace{-0.2in}
\end{table}

\begin{table}[t]
\caption{\textbf{\textit{Pixelation} effectiveness (extended).} FNR of four visual phishing detectors under \textit{Pixelation} attacks.} 
\label{tab:pixelation_fnr} 
% \vspace{-2mm}
\small
    \centering
    \resizebox{\linewidth}{!}{%
        \begin{tabular}{c|c|c|c|c|c}
            \toprule
            \textbf{Atk. intensity}&\textbf{Attack target}&\textbf{PhishIntention} & \textbf{PhishPedia} & \textbf{PhishLLM} & \textbf{VisualPhishNet}\\
            \hline
            \multirow{3}{*}{5x5px} & logo&1.0&1.0&0.278&0.833  \\ \cline{2-6}
           & background& 1.0&0.909 &0.364&0.818 \\ \cline{2-6}
           
           & logo \& background & 1.0&1.0&0.455&0.818   \\ \hline
            \multirow{3}{*}{4x4px} & logo&1.0&1.0&0.278&0.778 \\ \cline{2-6}
            
           & background& 1.0&0.909 &0.364&0.818  \\ \cline{2-6}
           & logo \& background & 1.0&1.0&0.455&0.818  \\ \hline
            \multirow{3}{*}{3x3px} & logo& 1.0&1.0&0.222&0.778 \\ \cline{2-6}
            
           & background&1.0&0.909 &0.545&0.818  \\ \cline{2-6}
           & logo \& background & 1.0&1.0&0.273&0.818 \\ \hline         \multirow{3}{*}{2x2px} & logo&1.0&0.944&0.222&0.778 \\ \cline{2-6}
           
           & background& 1.0&0.909 &0.273&0.818 \\ \cline{2-6}
           & logo \& background & 1.0&0.909&0.273&0.818  \\ \hline
           no-atk&baseline& 1.0&0.944&0.278&0.778  \\
           % \cline{2-6} 
            \bottomrule
        \end{tabular} 
        }

\end{table}

\begin{table}[t]
\caption{\textbf{Combination attacks effectiveness (extended).} FNR of four visual phishing detectors  under combined attacks (i.e., apply \textit{Pixelation} and \textit{Curtain effect} attacks simultaneously).} 
\label{tab:combine_fnr} 
% \vspace{-3mm}
\small
    \centering
    \resizebox{\linewidth}{!}{%
        \begin{tabular}{c|c|c|c|c|c}
            \toprule
            \textbf{Atk. intensity}&\textbf{Attack target}&\textbf{PhishIntention} & \textbf{PhishPedia} & \textbf{PhishLLM} & \textbf{VisualPhishNet}\\
            \hline
            \multirow{3}{*}{5x5px \& 25\%} &
            logo& 1.0&1.0&0.722&0.556 \\ \cline{2-6}
           & background& 1.0&0.909&0.455&0.636  \\ \cline{2-6}
           & logo \& background & 1.0&1.0&0.818&0.273  \\ \hline
           \multirow{3}{*}{5x5px \& 50\%} &
            logo& 1.0&1.0&0.333&0.722 \\ \cline{2-6}
           & background& 1.0&0.909&0.182&0.636 \\ \cline{2-6}
           & logo \& background & 1.0&1.0&0.455&0.636 \\ \hline
            \multirow{3}{*}{5x5px \& 75\%} &
            logo& 1.0&1.0&0.333&0.722 \\ \cline{2-6}
           & background&1.0&0.909&0.364&0.727   \\ \cline{2-6}
           & logo \& background & 1.0&1.0&0.273&0.636 \\ \hline
            \multirow{3}{*}{4x4px \& 25\%} &
            logo& 1.0&1.0&0.722&0.556 \\ \cline{2-6}
           & background& 1.0&0.909&0.364&0.636 \\ \cline{2-6}
           & logo \& background & 1.0&1.0&0.727&0.273 \\ \hline
           \multirow{3}{*}{4x4px \& 50\%} &
            logo& 1.0&1.0&0.389&0.722 \\ \cline{2-6}
           & background&1.0&0.909&0.182&0.636  \\ \cline{2-6}
           & logo \& background & 1.0&1.0&0.364&0.636  \\ \hline
            \multirow{3}{*}{4x4px \& 75\%} &
            logo&1.0&1.0&0.389&0.667  \\ \cline{2-6}
           & background& 1.0&0.909&0.455&0.727 \\ \cline{2-6}
           & logo \& background &1.0&1.0&0.273&0.636  \\ \hline

          \multirow{3}{*}{3x3px \& 25\%} &
            logo&1.0&1.0&0.722 &0.556  \\ \cline{2-6}
           & background& 1.0&0.909&0.364&0.636  \\ \cline{2-6}
           & logo \& background & 1.0&1.0&0.727&0.273 \\ \hline
           \multirow{3}{*}{3x3px \& 50\%} &
            logo& 1.0&1.0&0.333&0.722 \\ \cline{2-6}
           & background&1.0&0.909&0.182&0.636   \\ \cline{2-6}
           & logo \& background & 1.0&1.0&0.273&0.636 \\ \hline
            \multirow{3}{*}{3x3px \& 75\%} &
            logo& 1.0&1.0&0.389&0.667\\ \cline{2-6}
           & background& 1.0&0.909&0.364&0.727  \\ \cline{2-6}
           & logo \& background & 1.0&1.0&0.091&0.636 \\ \hline
     
            \multirow{3}{*}{2x2px \& 25\%} &
            logo& 1.0&1.0&0.778&0.556 \\ \cline{2-6}
           & background& 1.0&0.909&0.364&0.636 \\ \cline{2-6}
           & logo \& background & 1.0&1.0&0.818&0.273 \\ \hline
           \multirow{3}{*}{2x2px \& 50\%} &
            logo& 1.0&1.0&0.333&0.722\\ \cline{2-6}
           & background& 1.0&0.909&0.364&0.636 \\ \cline{2-6}
           & logo \& background & 1.0&1.0&0.455&0.636 \\ \hline
            \multirow{3}{*}{2x2px \& 75\%} &
            logo& 1.0&0.944&0.278&0.667\\ \cline{2-6}
           & background& 1.0&0.909&0.273&0.727  \\ \cline{2-6}
           & logo \& background &1.0&0.909&0.455&0.636  \\ \hline
        no-atk&baseline& 1.0&0.944&0.278&0.778 \\
            \bottomrule
        \end{tabular} 
        }
        
\end{table}